\documentclass[sn-mathphys-num]{sn-jnl}% Math and Physical Sciences Numbered Reference Style 
\usepackage{graphicx}%
\usepackage{multirow}%
\usepackage{amsmath,amssymb,amsfonts}%
\usepackage{amsthm}%
\usepackage{mathrsfs}%
\usepackage[title]{appendix}%
\usepackage{xcolor}%
\usepackage{textcomp}%
\usepackage{manyfoot}%
\usepackage{booktabs}%
\usepackage{algorithm}%
\usepackage{algorithmicx}%
\usepackage{algpseudocode}%
\usepackage{listings}%
\graphicspath{{./}{figures/}}

\setcitestyle{sort&compress,open={},close={},numbers,super}

\makeatletter 
\renewcommand\@biblabel[1]{#1} 
\makeatother

\usepackage{booktabs}\let\cline\cmidrule
\usepackage[font=scriptsize]{caption}
\usepackage[11pt]{extsizes}
\usepackage{geometry}
\usepackage{pdflscape}
\usepackage{rotating}
\usepackage[caption=false]{subfig}

\newcommand{\kms}{km s$^{-1}$}

\newcommand{\lya}{Ly$\alpha$}

\theoremstyle{thmstyleone}%
\theoremstyle{thmstyletwo}%

\theoremstyle{thmstylethree}%

\begin{document}

\title[Article Title]{Direct Images of the Cosmic Web of Intergalactic and
Circumgalactic Gas in the Distant Universe}


\author*[1]{\fnm{Kenneth M.} \sur{Lanzetta}}\email{kenneth.lanzetta@stonybrook.edu}

\author[2]{\fnm{Stefan} \sur{Gromoll}}

\author[3]{\fnm{Michael M.} \sur{Shara}}

\author[4]{Oleksii Sololiuk}

\author[5]{\fnm{David} \sur{Valls-Gabaud}}

\author[1]{\fnm{Anja} \sur{von der Linden}}

\author[1]{\fnm{Frederick M.} \sur{Walter}}

\author[6]{\fnm{John K.} \sur{Webb}}

\affil*[1]{\orgdiv{Department of Physics and Astronomy}, \orgname{Stony Brook University}, \orgaddress{\city{Stony Brook}, \state{NY}, \postcode{11794-3800},
\country{USA}}}

\affil[2]{\orgname{Amazon Web Services}, \orgaddress{\street{410 Terry Ave.\ N}, \city{Seattle}, \state{WA}, \postcode{98109}, \country{USA}}}

\affil[3]{\orgdiv{Department of Astrophysics}, \orgname{American Museum of Natural History}, \orgaddress{\street{Central Park West at 79th St.}, \city{New York}, \state{NY}, \postcode{10024-5192}, \country{USA}}}

\affil[4]{Department of Physics, University of Aberdeen, Aberdeen AB24 3UE, UK}

\affil[5]{\orgdiv{LERMA, CNRS}, \orgname{Observatoire de Paris}, \orgaddress{\street{61 Avenue de l'Observatoire}, \city{Paris}, \postcode{75014}, \country{FRANCE}}}

\affil[6]{\orgdiv{Institute of Astronomy}, \orgname{University of Cambridge}, \orgaddress{\street{Madingley Road}, \city{Cambridge}, \postcode{CB3 0HA}, \country{UK}}}

\maketitle

{\large Most of the baryonic matter of the Universe resides in a highly-ionized
gaseous intergalactic medium\cite{cen2006}.  This gas flows along dark-matter
filaments toward galaxy superclusters, clusters, and groups until it pools
around the galaxies into a circumgalactic medium\cite{tum2017}.  Eventually,
the gas settles into the interstellar medium of the galaxies, where it fuels
the successive generations of star formation that ultimately produce the stars
and heavy elements that make up galaxies today.  The gas has been studied for
decades using absorption lines produced by Hydrogen and various ions of heavy
elements in the spectra of background quasi-stellar objects (QSOs).  But
directly imaging the extremely faint glow of this ``cosmic web'' of
intergalactic and circumgalactic gas has remained an elusive goal of
observational cosmology.  Some recent progress has been made by using imaging
spectrographs to record high-redshift \lya\ emission, although over only very
narrow fields of view\cite{can2014,gal2018,bac2021,mar2023}.  Here we report
direct images of intergalactic and circumgalactic gas in the distant Universe
obtained using the Condor Array Telescope that we purposely built to reach
extremely low-surface-brightness sensitivities over very wide fields of
view\cite{lan2023}.  We show that these images directly detect and
characterize the imprint of \lya\ emission from the cosmic web at an
overwhelming statistical significance.  By stacking portions of the images
centered on tens of thousands of galaxies of known redshift, we show that they
also reveal extremely faint emission from H$^0$, C$^{3+}$, and Mg$^+$ and
absorption from cosmic dust in the tenuous outskirts of the galaxies.  Our
results demonstrate that sensitive imaging observations can now detect and
characterize emission (and absorption) from the cosmic web of intergalactic and
circumgalactic gas (and dust).}

Almost everything that is known about extremely tenuous gas in the distant
Universe has been learned by studying absorption lines produced by foreground
objects in the spectra of background QSOs.  It became apparent soon after the
discovery of the first QSO 3C 273 in 1963\cite{haz1963, sch1963} that narrow
absorption lines are ubiquitous in QSO spectra and that the lines are produced
by cosmologically distributed gas intervening along the lines of
sight\cite{lyn1971}.  The origin of the gas has been surmised to include
objects ranging from the extended ``halos'' of normal galaxies\cite{bah1969}
through intergalactic clouds\cite{sar1980} through more recently the cosmic web
that is predicted by hydrodynamical simulations of large-scale structure
formation\cite{sch2015,bol2017}.  Neutral Hydrogen column densities of the gas 
range from $\approx 10^{12}$ to $\approx 10^{17}$ cm$^{-2}$ for the
``\lya-forest'' absorbers through $\approx 10^{17}$ to $\approx 10^{20}$
cm$^{-2}$ for the ``Lyman-limit'' absorbers through $\approx 10^{20}$ to
$\approx 10^{22}$ cm$^{-2}$ for the ``damped-\lya'' absorbers.

The prospect of detecting \lya\ emission from the gas responsible for QSO
absorption lines was first contemplated in the 1980s\cite{hog1987}.  The
difficulty is that the emission is predicted to be extremely faint, with the
brightness of gas fluorescing due to the cosmic ultraviolet radiation field at
redshift $z \approx 3$ expected to be only $\approx 10^{-20}$ erg s$^{-1}$
cm$^{-2}$ arcsec$^{-2}$ in \lya\cite{can2005}.  This is fainter than the
brightness of the night sky by a factor $\approx 1000$ over even a very narrow
bandpass.  Recent efforts to detect this emission have used the Multi Unit
Spectral Explorer (MUSE) on the Very Large Telescope (VLT)\cite{bac2021} at
redshifts $z \approx 3.8$ and the Cosmic Web Imager (CWI) on the Keck
telescope\cite{mar2023} at redshifts $z \approx 2.35$.  These instruments are
imaging spectrographs that provide substantial wavelength coverage (e.g.\
$48.5$ nm for CWI) over very narrow fields of view (e.g.\ $30 \times 20$
arcsec$^2$ for CWI).  Accordingly, these instruments obtain ``core samples''
that extend much farther in the radial direction than in the transverse
direction.  For example, each CWI exposure extends (for \lya) $\approx 900$
comoving Mpc in the radial direction but only $\approx 0.86 \times 0.58 = 0.50$
Mpc$^2$ in the transverse direction, for a total volume of $\approx 900 \times
0.86 \times 0.58 \approx 450$ Mpc$^3$, with similar values applying for MUSE
exposures.  (We adopt a $\Lambda$CDM cosmological model of matter density
parameter $\Omega_m = 0.3$, vacuum energy density parameter $\Omega_\Lambda =
0.7$, and Hubble constant $70$ km s$^{-1}$ Mpc$^{-1}$ throughout.)

We used the Condor Array Telescope\cite{lan2023} to obtain very deep direct
images of Condor field 416, which encompasses and extends most of the Cosmic
Evolution Survey (COSMOS) field\cite{sco2013}, through luminance broad-band and
custom narrow-band filters.  The luminance filters are sensitive to wavelengths
$\approx 400$ to $700$ nm.  The narrow-band filters are tuned to a central
wavelength (in the Condor $f/5$ beam) of $422.5$ nm with a bandpass of only $1$
nm (see Extended Data Fig.\ 1).  The central wavelength corresponds to a
redshift $z = 2.4754$ for \lya, which was chosen based on results of
\setcitestyle{sort&compress,open={},close={},numbers}
\hspace{-1.1em}
ref.\ \citep{mar2023}
\hspace{-1.1em}
\setcitestyle{sort&compress,open={},close={},numbers,super}
(see below) and is proximate to the Hyperion protocluster at redshift $z
\approx 2.45$\citep{cuc2018}.  The bandpass corresponds to a velocity interval
of $\Delta v \approx 710$ \kms\ and a redshift interval of $\Delta z \approx
0.0059$ for \lya.  The exposure time through the luminance filters totals 26.3
h and through the narrow-band filters totals 171.8 h, where both values are
equivalent exposure times for the full array.  The images were processed using
the Condor data pipeline\citep{lan2023, lan2023a} (see Methods).  The final
processed narrow-band image reaches a $1 \sigma$ uncertainty of $3 \times
10^{-19}$ erg s$^{-1}$ cm$^{-2}$ arcsec$^{-2}$ per $0.85 \times 0.85$
arcsec$^{-2}$ pixel, which corresponds to a $3 \sigma$ uncertainty of $8 \times
10^{-20}$ erg s$^{-1}$ cm$^{-2}$ arcsec$^{-2}$ per $10 \times 10$ arcsec$^2$
region.  This is essentially {\em identical} to the sensitivity of the CWI
image of the COSMOS field by
\setcitestyle{sort&compress,open={},close={},numbers}
\hspace{-0.8em}
ref.\ \citep{mar2023},
\hspace{-0.8em}
\setcitestyle{sort&compress,open={},close={},numbers,super}
which reaches a $3 \sigma$ depth for a 0.8-nm-wide line (i.e.\ comparable to
the Condor 1 nm bandpass) of $8 \times 10^{-20}$ erg s$^{-1}$ cm$^{-2}$
arcsec$^{-2}$ per $10 \times 10$ arcsec$^2$ region.  The corresponding
values for the final processed luminance image are a $1 \sigma$ uncertainty of
$4.8 \times 10^{-32}$ erg s$^{-1}$ cm$^{-2}$ Hz$^{-1}$ arcsec$^{-2}$ (or $29.7$
mag arcsec$^{-2}$) per pixel and a $3 \sigma$ uncertainty of $1.4 \times
10^{-32}$ erg s$^{-1}$ cm$^{-2}$ Hz$^{-1}$ arcsec$^{-2}$ (or $31.0$ mag
arcsec$^{-2}$) per $10 \times 10$ arcsec$^{-2}$ region.

The Condor field of view is $\approx 2.3 \times 1.5$ deg$^2$, and each Condor
narrow-band exposure extends (for \lya) $\approx 18$ comoving Mpc in the radial
direction and $\approx 240 \times 160 \approx 38,000$ Mpc$^2$ in the transverse
direction, which exceeds the CWI value by a factor $\approx 76,000$.  The total
volume probed by each Condor exposure is $\approx 18 \times 240 \times 160
\approx 690,000$ Mpc$^3$, which exceeds the CWI value by a
\begin{landscape}
\begin{figure}
\centering
\vspace{-0.25in}
\includegraphics[width=1.00\linewidth, angle=0, viewport=97 62 745 530]{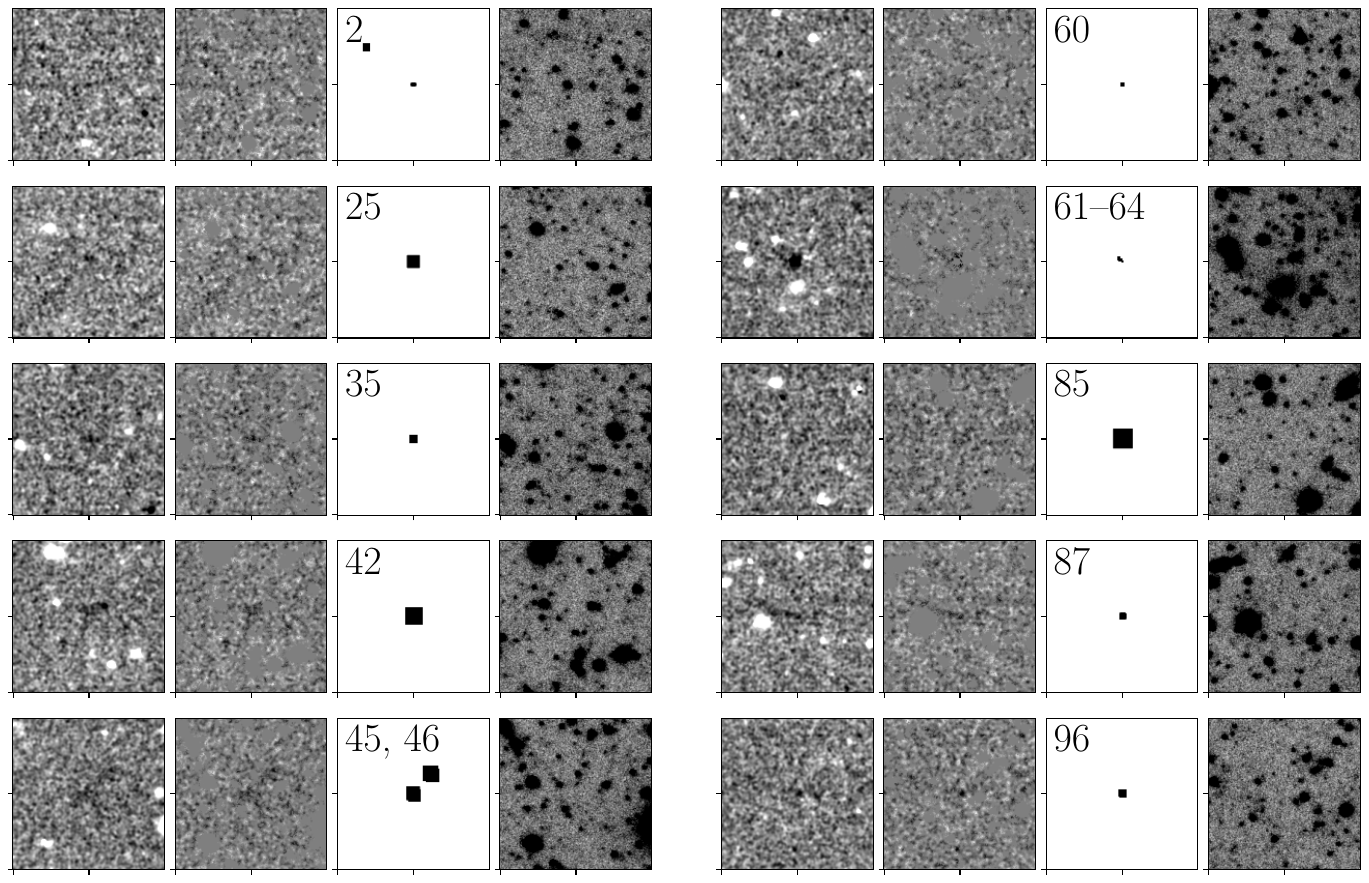}
\caption{Examples of detected features of (1) high statistical significance and
(2) large rest-frame equivalent width (for \lya).  Each panel shows small
portions of (left to right) difference image, masked difference image,
detection map, and luminance image.  Images extend $2.8 \times 2.8$ armin$^2$,
which at redshift $z = 2.4754$ corresponds to $1.4$ proper Mpc or $4.9$ Mpc
comoving Mpc on a side.  Brightest single pixels in portions of masked
difference image are of surface brightness $\approx 10^{-18}$ erg s$^{-1}$
cm$^{-2}$ arcsec$^{-2}$.}
\end{figure}
\end{landscape}
{\flushleft factor $\approx 1500$.  The ability of Condor to probe enormous
transverse areas and volumes one redshift slice at a time gives it capabilities
that are complementary to those of MUSE and CWI.}

The luminance image shows evidence of modest Galactic cirrus toward the SE edge
of the image, around J2000 coordinates 10:07, $+$02:15, which does not appear
to overlap the COSMOS field proper.  To at least a first approximation, any
line emission at $422.5$ nm is insignificant in comparison to continuum over
the very large bandpass of the luminance filter, so the luminance image roughly
traces continuum, whereas the narrow-band image traces line emission plus
continuum.  We subtracted the luminance image from the narrow-band image to
form a ``difference'' image that in most regions traces more or less only line
emission, thus accounting for any extremely faint cirrus that may be present
across the field.  (The difference image does not necessarily trace only line
emission in regions near bright stars and galaxies, for which continuum color
effects can be significant.)  The narrow-band, luminance, and difference images
are shown in Extended Data Fig.\ 2, 3, and 4, respectively.

Low-surface-brightness \lya\ emission of redshift $z \approx 2.4754$ from the
cosmic web could in principle be seen only in regions between continuum sources
and would be characterized by large rest-frame equivalent width limits (say
$\gtrsim 5$ nm), i.e.\ would appear much brighter in the narrow-band image than
in the luminance image.  Such emission {\em is} seen throughout the entire
field of view.  Detection of this faint emission may be optimized by processing
the images with a matched filter\cite{tur1960}, which requires knowledge of the
sizes and shapes of the emitting features.  Lacking such knowledge, we searched
for faint emission by binning the images on scales ranging up to $50 \times 50$
pix$^2$, seeking to identify features of (1) high statistical significance and
(2) large rest-frame equivalent width (see Methods).  This analysis, which is
most sensitive to compact and concentrated (rather than irregular or
filamentary) sources, identified 112 features of statistically-significant,
strong line emission that can be reasonably interpreted only as \lya\ of
redshift $z \approx 2.4754$.  The resulting catalog of features is presented in
Extended Data Table 1, and images of the features are presented in Extended
Data Figure 5.  The narrow-band image is also sensitive to [O~II] 372.7 nm
emission of redshift $z \approx 0.1336$, and indeed such emission is seen, but
of much lower rest-frame equivalent width.

Examples of several of the detected features are illustrated in Fig.\ 1, which
shows small portions of the difference image, the ``masked'' difference image
(which was formed by masking pixels of the difference image corresponding to
pixels detected in the luminance image above a $1 \sigma$ threshold), the
detection map, and the luminance image.  Features shown in Fig.\ 1 include a
compact source (35); compact sources that appear to be embedded in filamentary
structures that stretch up to and beyond $1$ Mpc in extent (2, 25, 60, 87, and
96); diffuse, extended sources (42, 45, 46, and 85); and emission in the
vicinity of a known active galactic nucleus (AGN) ACS--GC
20118705\cite{tru2009} (61 through 64, and see below).  The brightest single
pixels in the portions of the masked difference image (i.e.\ that are free of
continuum) shown in Fig.\ 1 are of surface brightness $\approx 10^{-18}$ erg
s$^{-1}$ cm$^{-2}$ arcsec$^{-2}$.

The distribution of pixel intensities of the difference image in regions
between continuum sources is sensitive to the surface-brightness intensity
distribution\cite{lan2002} of \lya\ emission at redshift $z \approx 2.4754$
from the cosmic web, irrespective of whether emission is or is not ``detected''
in any given pixel.  This is illustrated in Fig.\ 2, which shows distributions
of pixel intensities of the masked difference image (see Methods).  The top
panel of Fig.\ 2 shows the observed distribution together with a $\chi^2$ fit
to a normal distribution.  The observed distribution appears to be very well
described by a normal distribution except for an apparent systematic excess of
pixels of positive intensity $\gtrsim 7 \times 10^{-19}$ erg s$^{-1}$ cm$^{-2}$
arcsec$^{-2}$.  That the pixel-to-pixel measurement uncertainties would be
characterized by a normal distribution is not unexpected given that the
narrow-band image is formed by combining 6185 dithered individual exposures.
The fit yields $\chi^2 = 562.9$ for 129 degrees of freedom, which indicates a
statistically unacceptable fit, with a $p$ value of essentially zero.  We
interpret the departures of the observed distribution from a normal
distribution as due to \lya\ emission from the cosmic web.

\begin{figure}
\centering
\includegraphics[width=1.00\linewidth, angle=0]{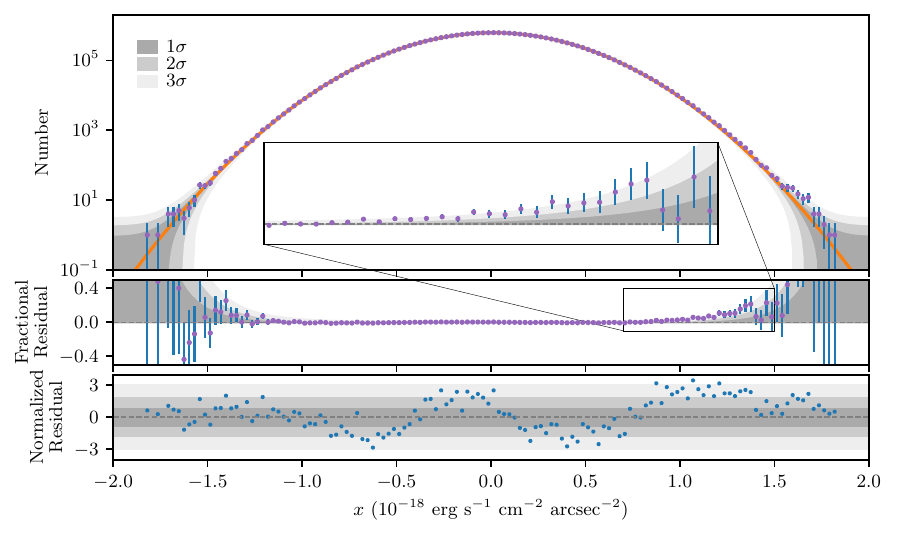}
\includegraphics[width=1.00\linewidth, angle=0]{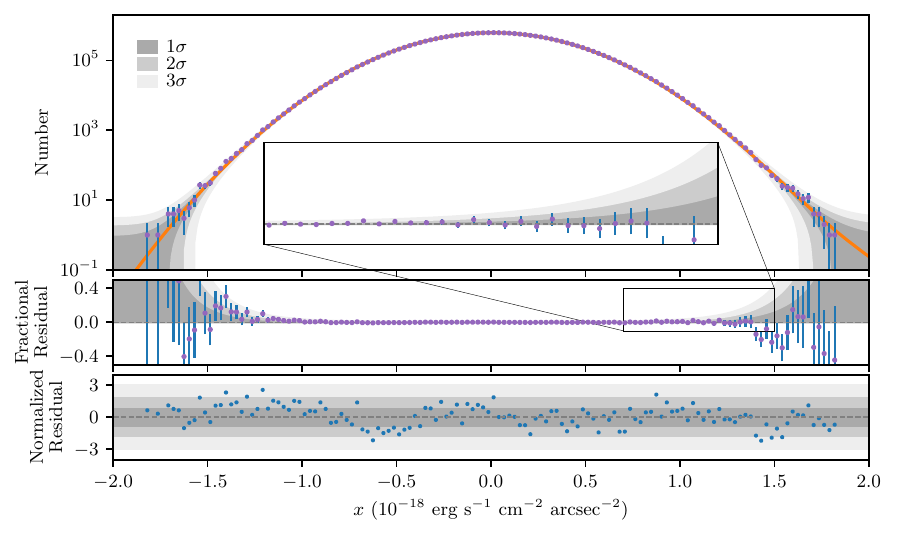}
\caption{Distributions of pixel intensities of masked difference image, i.e.\ of
pixels of difference image between continuum sources.  Top panel
shows observed distribution fitted by normal distribution of best-fit parameter
values $\sigma = 3.3878 \pm 0.0021 \times 10^{-19}$ erg s$^{-1}$ cm$^{-2}$
arcsec$^{-2}$ and $\bar{x} = 1.166 \pm 0.014 \times 10^{-20}$ erg s$^{-1}$
cm$^{-2}$ arcsec$^{-2}$.  Bottom panel shows observed distribution fitted by
model of equation (2) adopting $x_{\rm min} = 10^{-21}$ erg s$^{-1}$ cm$^{-2}$
arcsec$^{-2}$ (although fit is insensitive to choice of $x_{\rm min}$) of
best-fit parameter values $\sigma = 3.3727 \pm 0.0021 \times 10^{-19}$ erg
s$^{-1}$ cm$^{-2}$ arcsec$^{-2}$, $\bar{x} = 5.65 \pm 0.14 \times 10^{-21}$ erg
s$^{-1}$ cm$^{-2}$ arcsec$^{-2}$, $\alpha = 0.66 \pm 0.50$, $x_0 = 1.286 \pm
0.035 \times 10^{-19}$ erg s$^{-1}$ cm$^{-2}$ arcsec$^{-2}$, and $B/A = 0.32
\pm 0.20$.  In each case, top plot shows observed and fitted distributions,
middle plots show fractional residuals, and bottom plot show normalized
residuals (in units of standard deviation $\sigma$), and inset shows zoom of
portion of middle plot.  Grey regions show 1-, 2-, and 3-$\sigma$ confidence
intervals.}
\end{figure}

Motivated by the intensity distribution of the rest-frame ultraviolet continuum
of high-redshift galaxies\cite{lan2002} and the H$^0$ column density
distribution of QSO absorption lines\cite{pet1993}, we choose to describe the
\lya\ surface brightness distribution $h(x)$ as a function of \lya\
surface-brightness intensity $x$ by a truncated power law with exponential
cutoff
\begin{equation}
h(x)= A \delta(x) + \frac{B}{x_0} H(x - x_{\rm min}) \left( \frac{x}{x_0}
\right)^{-\alpha} \exp{\left( -\frac{x}{x_0} \right)},
\end{equation}
where $\delta(x)$ is the Dirac delta function and $H(x)$ is the Heaviside step
function.  Here the delta function term and the faint-end truncation at
intensity $x_{\rm min}$ serve to regularize the distribution (see Methods),
$x_0$ is a characteristic intensity, $\alpha$ is a power-law index, and the
ratio $B/A$ is related to the fraction of the sky covered by emission.  The
observed distribution $g(x)$ is the convolution of $h(x)$ with the
pixel-to-pixel measurement uncertainty distribution, which we take to be a
normal distribution $N(x; \sigma, \bar{x})$ of standard deviation $\sigma$ and
mean $\bar{x}$, i.e.
\begin{equation}
\begin{split}
g(x; \alpha, x_0, B/A, \sigma, \bar{x}) & = \int_0^\infty
h(x'; \alpha, x_0, B/A) N(x - x'; \sigma, \bar{x}) \ dx' \\
& = A N(x; \sigma, \bar{x} ) + \frac{B}{x_0} \int_{x_{\rm min}}^\infty
\left( \frac{x'}{x_0} \right)^{-\alpha} \exp{ \left( -\frac{x'}{x_0} \right) }
N(x - x'; \sigma, \bar{x}) \ dx'.
\end{split}
\end{equation}
(A non-zero value of $\bar{x}$ can account for small errors in background
determination.)  The bottom panel of Fig.\ 2 shows the observed distribution
together with a $\chi^2$ fit to the model of equation (2).  The fit yields
$\chi^2 = 139.2$ for 126 degrees of freedom, which indicates a statistically
acceptable fit, with a $p$ value of 0.20.  Thus the fit to the model of
equation (2) is statistically superior to the fit to a normal distribution,
reducing $\chi^2$ by $\Delta \chi^2 = 423.7$ with the introduction of only
three additional parameters; this is extremely unlikely by chance, with a $p$
value of essentially zero.  We take the difference between the top and bottom
panels of Fig.\ 2 and in particular the difference between the goodness of fit
of the respective fits as demonstrating that the difference image in regions
between continuum sources detects the imprint of \lya\ emission from the cosmic
web at redshift $z \approx 2.4754$ at an overwhelming statistical significance.

The \lya\ surface brightness intensity distribution $h(x)$ is a fundamental
statistical description of the cosmic web that must be matched by any
successful cosmological simulation of large-scale structure formation.
Further, the first moment of the distribution is the \lya\ luminosity density.
Adopting the best-fit parameter values of the model of equation (2) given in
the caption to Fig.\ 2, the comoving \lya\ luminosity density of the cosmic web
at redshift $z \approx 2.4754$ is $2.9 \pm^{2} \times 10^{40}$ erg s$^{-1}$
Mpc$^{-3}$.  We emphasize that this value includes contributions from {\em all}
sources of \lya\ emission in the difference image in regions between continuum
sources, including \lya-emitting galaxies, the circumgalactic medium, and the
intergalactic medium.

The central wavelength of the narrow-band filter was chosen to correspond to
the redshift $z = 2.4754$ of a $1 \times 1$ arcmin$^2$ region of apparent
particularly high density of \lya\ emission in the CWI image presented in the
middle, right-most panel of Fig.\ 4b of
\setcitestyle{sort&compress,open={},close={},numbers}
\hspace{-1.0em}
ref.\ \citep{mar2023}.
\hspace{-1.0em}
\setcitestyle{sort&compress,open={},close={},numbers,super}
Portions of the smoothed difference, difference, and luminance images
containing this region are presented in Fig.\ 3, with the boundary of the
CWI image indicated.  Both the Condor and CWI images show obvious \lya\
emission of redshift $z \approx 2.4754$ from the AGN ACS--GC
20118705\cite{tru2009} (near the SW corner of the bounded region), but
otherwise there is little correspondence between the images.  The Condor image
shows features not seen in the CWI image, and vice versa.  In particular, the
Condor image shows two apparent filaments or tails in the vicinity of
ACS--GC 20118705 that are not seen in the CWI image, one at roughly 2 o'clock
and one at roughly 10 o'clock; the Condor image also shows various
spatially-unresolved sources not seen in the CWI image, and vice versa.
(Indeed, sources 61 through 64 of Extended Data Table 1 and Extended Data Fig.\
5 are detected in the vicinity of the AGN.)  We cannot explain this difference,
although we note that the Condor difference image is a difference between two
direct images whereas the CWI image is a highly-processed product that involves
iterative adaptive smoothing of spectroscopic observations.

\begin{figure}
\centering
\includegraphics[width=0.99\linewidth, angle=0]{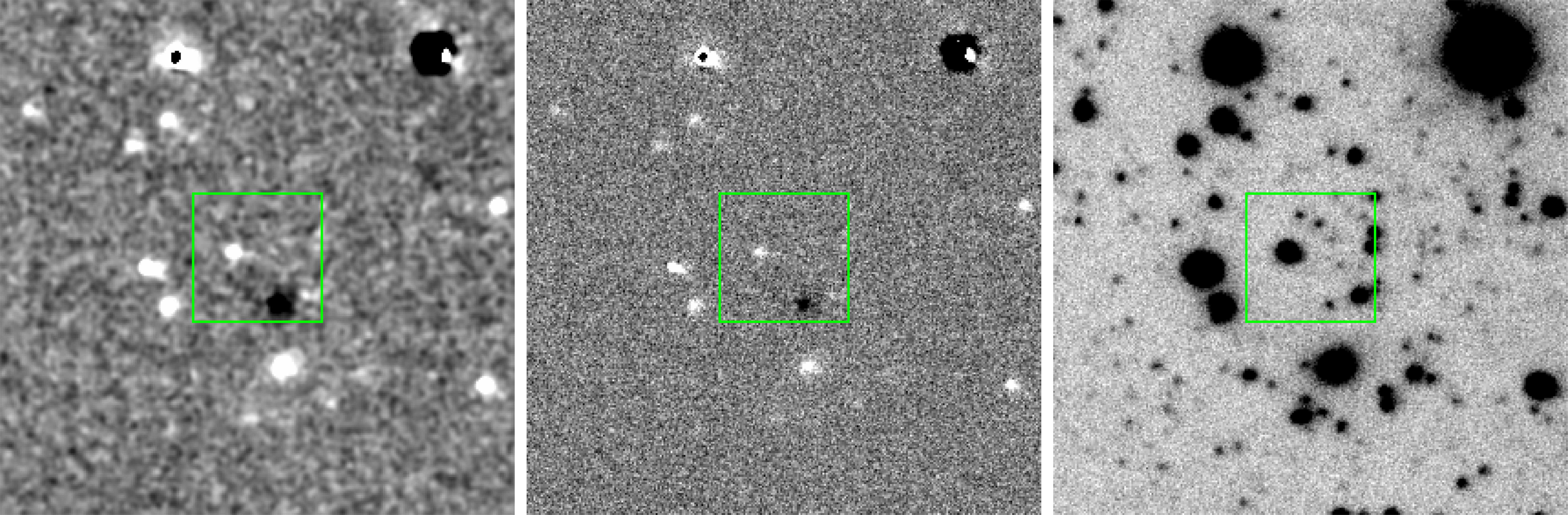}
\setcitestyle{sort&compress,open={},close={},numbers}
\caption{Portions of (left to right) smoothed difference, difference, and
luminance images containing $1 \times 1$ arcmin$^2$ region (shown by green box)
of apparent particularly high density of \lya\ emission in CWI observations.
Region within green box may be directly compared with middle, right-most panel
of Fig. 4b of ref.\ \citep{mar2023}.  Images extend $4 \times 4$ arcmin$^2$
centered at J2000 coordinates 10:00:22, $+$02:24:18, which at redshift $z =
2.4754$ corresponds to $2.0$ proper Mpc or $7.0$ comoving Mpc on a side.
Smoothed difference image is smoothed by Gaussian of standard deviation 1.5
pix.  North is up, and east is to left.}
\setcitestyle{sort&compress,open={},close={},numbers,super}
\end{figure}

Nearly two million galaxy redshifts have been measured in the COSMOS field (by
spectroscopic or photometric techniques), which provides an opportunity to
``stack'' cutouts of images of many galaxies of similar redshift in order to
form a ``composite'' image of the galaxies that is far more sensitive than any
single image alone.  We formed stacked median cutouts of the luminance image of
galaxies over a range of redshift and brightness (see Methods).  An example of
the result for galaxies of redshift near the primary target redshift $z =
2.4754$ is presented in Fig.\ 4, for galaxies in four brightness ranges.  For
each brightness range, Fig.\ 4 shows the composite image and the radial profile
measured from the center of the image.  In each case, the composite image
exhibits (1) a bright core, which is produced by starlight emitted by the
galaxies, and (2) a dark ``halo'' surrounding the core that extends $\approx
250$ proper kpc in radius.  The composite images of all but the brightest
galaxies further exhibit (3) an extended dark halo that extends at least
$\approx 1.7$ proper Mpc or $\approx 6.0$ comoving Mpc.  The composite images
are approximately azimuthally symmetric, as is expected if the galaxies are
randomly distributed with no preferred orientation.

The dark halos and extended dark halos appear to be universal and are present
in composite images of galaxies of redshift spanning $z = 0$ through beyond 3
and a range of brightnesses, although there is a tendency for the halos of the
brightest galaxies at a given redshift to be less pronounced.  We found that
the effect persists across different stacking algorithms (including median and
sigma clipping), is apparent in stacks that we formed from the luminance and
narrow-band images, and is even apparent in stacks that we formed from Subaru
broad-band images of the COSMOS field\cite{tan2015}.  We conclude that the
effect is {\em not} the result of an instrumental or processing artifact.

We interpret the dark halos surrounding the bright cores as due to obscuration
of background galaxies by the dusty circumgalactic medium of the galaxies.
This sets a characteristic scale of $\approx 250$ proper kpc radius for the
circumgalactic medium in dust at redshift $z \approx 2.48$, which is in line
with other measures of extended dust around galaxies\cite{yor2006,men2010}.  We
consider the composite images of Fig.\ 4 (and similar images formed at other
redshifts and brightnesses) to be vivid and graphic direct images of the
circumgalactic medium of galaxies in dust.  We speculate that perhaps the
effect is less apparent in brighter galaxies because brighter galaxies are
preferentially early-type galaxies whereas fainter galaxies are preferentially
late-type galaxies, and a dusty circumgalactic medium may be less prevalent
around early-type galaxies.

\begin{figure}
\centering
\includegraphics[width=1.00\linewidth, angle=0]{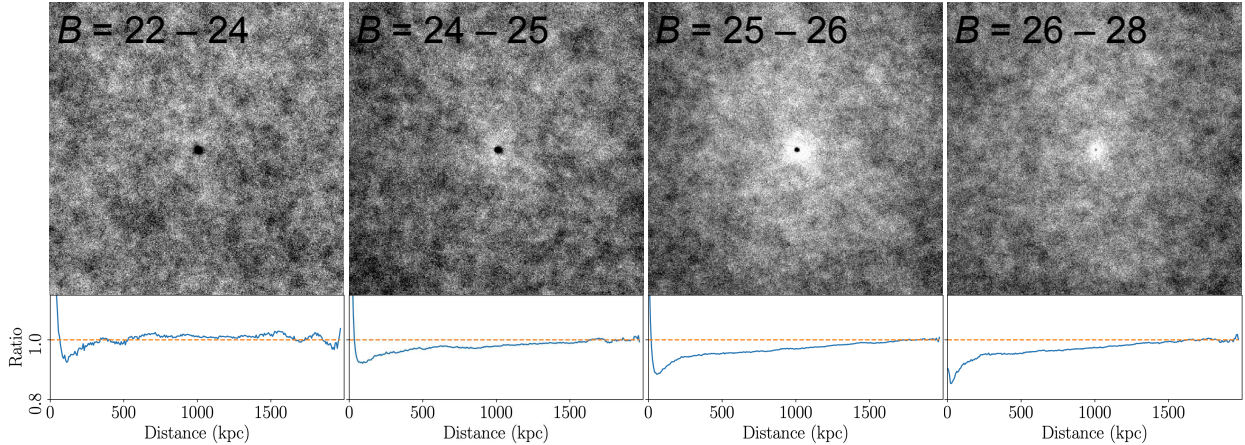}
\caption{Stacked median cutouts of luminance image (top panels) and radial
profiles (bottom panels) of galaxies of redshift within $\Delta z = 0.5$ of $z
= 2.5$ and of (left to right) $B$-band magnitude 22 to 24 (554 galaxies), 24 to
25 (4355 galaxies), 25 to 26 (8828 galaxies), and 26 to 28 (4877 galaxies).
Radial profiles are normalized to unity at right-hand edges.  Images extend
$5.7 \times 5.7$ arcmin$^2$, which corresponds to  2.8
proper Mpc or 9.7 comoving Mpc on a side.  Images are reversed, with emission
dark.}
\end{figure}

The extended dark halos surrounding the bright cores are far too large to be
associated with individual galaxies and indeed extend to scales comparable to
the galaxy correlation length of $\approx 7$ comoving Mpc\cite{pee1975}.  We
interpret the extended dark halos as due to obscuration of background galaxies
by dusty material spread on intergalactic scales, i.e.\ overlapping dusty halos
of correlated galaxies or dust distributed in intergalactic space or both.
Future analyses will more completely characterize properties of cosmic dust in
the circumgalactic and perhaps intergalactic medium.

A similar stacking technique can be applied to the narrow-band image to search
for faint fluorescence emission from ions distributed in the extended gaseous
envelopes of galaxies that are inferred on the basis of statistical comparison
of QSO absorption lines and galaxies along the lines of sight in
\setcitestyle{sort&compress,open={},close={},numbers}
\hspace{-0.7em}
H$^0$~(ref.\ \citep{lan1995}), C$^{3+}$~(ref.\ \citep{che2001}), and
Mg$^+$~(ref.\ \citep{ber1991}).
\hspace{-0.5em}
\setcitestyle{sort&compress,open={},close={},numbers,super}
Here the stacked median cutouts of the narrow-band image involve a complex
blend of emission (including continuum emission from foreground and background
galaxies and continuum and line emission from the target galaxies) and
absorption (including line absorption by gas and continuum absorption by dust
of the target galaxies).  To address this complexity, we formed ``on-band'' and
``off-band'' stacked median cutouts of the narrow-band image of galaxies of
various redshifts (see Methods).  Redshifts of the galaxies included into the
on-band images shift a resonance transition into the filter bandpass, and
redshifts of the galaxies included into the off-band images do not shift any
resonance transition into the filter bandpass.  Subtracting the off-band image
from the on-band image then removes continuum emission and absorption, leaving
only the line emission and absorption of the resonance transition.  In
practice, we bracketed each on-band image with a pair of off-band images, one
formed from slightly lower redshifts and one formed from slightly higher
redshifts, and we subtracted the mean off-band image (thereby accounting for
the possibility of gradients in the continuum emission and absorption with
wavelength).  We used a weighted median stacking method to allow for galaxy
redshift uncertainties, with weights determined by taking the uncertainties to
be normally distributed with a standard deviation of $0.1$.

\begin{figure}
\centering
\includegraphics[width=1.00\linewidth, angle=0]{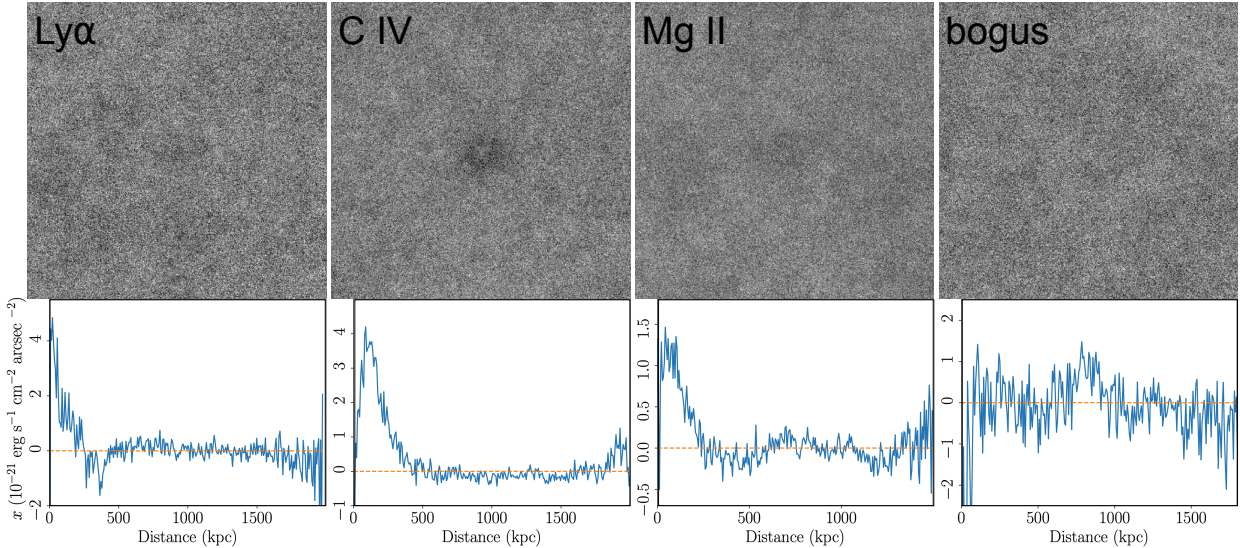}
\caption{Composite on-band minus off-band images (top panels) and radial
profiles (bottom panels) for resonance transitions of (left to right) \lya\
from galaxies of redshift $z \approx 2.48$ ($6935$ on-band galaxies), C~IV
154.9 nm from galaxies of redshift $z = 1.73$ ($24,100$ on-band galaxies), and
Mg~II 279.6 nm from galaxies of redshift $z \approx 0.51$ ($55,270$ on-band
galaxies) and for a ``bogus'' redshift $z = 3.08$ that does not correspond to
any resonance transition ($3499$ on-band galaxies).  Analysis incoporates
roughly one to three times as many off-band galaxies, depending on transition.
Images extend $5.7 \times 5.7$ arcmin$^2$, which corresponds to 2.1 to 2.9
proper Mpc or 3.2 to 10.8 comoving Mpc on a side, depending on redshift.
Images are reversed, with emission dark.}
\end{figure}

Results for the resonance transitions of \lya\ from galaxies of redshift $z
\approx 2.48$, C~IV 154.9 nm from galaxies of redshift $z = 1.73$, and Mg~II
279.6 nm from galaxies of redshift $z \approx 0.51$ and for a ``bogus''
redshift $z = 3.08$ that does not correspond to any resonance transition are
presented in Fig.\ 5.  For each transition at each redshift, Fig.\ 5 shows the
composite on-band minus off-band image and the radial profile measured from the
center of the image.  In each real case, the composite image shows extremely
faint but significant emission.  The composite image of \lya\ emission
indicates a peak median intensity of $\approx 4 \times 10^{-21}$ erg s$^{-1}$
cm$^{-2}$ arcsec$^{-2}$ and an extent of $\approx 250$ proper kpc, the
composite image of C~IV emission indicates a peak median intensity of $\approx
4 \times 10^{-21}$ erg s$^{-1}$ cm$^{-2}$ arcsec$^{-2}$ and an extent of
$\approx 400$ kpc, and the composite image of Mg~II emission indicates a peak
median intensity of $\approx 1 \times 10^{-21}$ erg s$^{-1}$ cm$^{-2}$
arcsec$^{-2}$ and an extent of $\approx 250$ kpc.  In the bogus case, the
composite image does not show significant emission.  The extent of the \lya\
emission of redshift $z \approx 2.48$ indicated by the composite image of the
first panel of Fig.\ 5 is in remarkable agreement with the extent $\approx 230$
kpc of the \lya\ absorption at redshifts $z \lesssim 1$ inferred by comparing
QSO absorption lines and galaxies\cite{lan1995}.  And, strikingly, the radial
profile of the C~IV image of Fig.\ 5 exhibits a clear ring-like structure,
which is consistent with the long-held notion that the extended envelopes of
C~IV-absorbing gas surround the extended envelopes of \lya-absorbing
gas\cite{che2001}.  We consider the composite images of Fig.\ 5 to be vivid and
graphic direct images of the circumgalactic medium of galaxies in gas.

Our results clearly demonstrate that Condor, equipped with broad- and
narrow-band filters and dedicated to obtaining very long exposures, can detect
and characterize the cosmic web in \lya\ and can detect and study the
circumgalactic medium in gas and dust.  Some, but not all, of our results rely
on the huge number of galaxy redshifts that have been measured in the COSMOS
field, and these results will be difficult to replicate or build upon in other
fields.  But the COSMOS field itself remains ripe for exploration, and future
observations and analyses of this and other fields will probe deeper into the
cosmic web and will establish properties of intergalactic and circumgalactic
gas and dust as functions of galaxy environment and cosmic epoch.

\section*{Methods}

\subsection*{Observations and Data Processing}

Condor is an ``array telescope'' that consists of six apochromatic refracting
telescopes of objective diameter 180 mm, each equipped with a large-format
($9576 \times 6388$ pix$^2$), very-low-read-noise ($\approx 1.2$ e$^-$),
very-rapid-read-time ($< 1$ s) CMOS camera\cite{lan2023}.  Over the period
stretching from January through June 2024, we used Condor to obtain very deep
direct images of Condor field 416 through luminance broad-band and custom
narrow-band filters.  The total exposure time of 26.3 h through the luminance
filters and 171.8 h through the narrow-band filters corresponds to a
reach\cite{lan2024a} of the luminance observations of $1.4 \times 10^4$ m$^2$ s
and of the narrow-band observations of $9.5 \times 10^4$ m$^2$ s.  All
individual exposures were obtained with an exposure time of 600 s, and the
telescope was dithered by a random offset of $\approx 15$ arcmin between each
individual exposure.

We worked together with Chroma Technology Corp.\ to design and fabricate the
custom narrow-band filters tuned to a central wavelength in the Condor $f/5$
beam of $422.5$ nm with a bandpass of only $1$ nm and a peak throughput of
$\approx 85\%$.  The filter response function (measured by Chroma) together
with an indication of the predicted central wavelength in the Condor $f/5$
beam (calculated by Chroma) is shown in Extended Data Fig.\ 1.

We processed the observations using the Condor data pipeline\cite{lan2023,
lan2023a}.  The data pipeline performs bias subtraction; field flattening and
background subtraction; astrometric calibration; identification and masking of
cosmic ray events, satellite trails, and pixels that exhibit significant random
telegraph noise; photometric calibration; drizzling onto a common coordinate
grid; and coaddition of the individual exposures.  The narrow-band, luminance,
and difference images are shown in Extended Data Fig.\ 2, 3, and 4
respectively.

\subsection*{Detection of Faint Emission}

To detect low-surface-brightness \lya\ emission of redshift $z \approx 2.4754$
in regions between continuum sources, we binned the narrow-band and luminance
images on scales ranging from $1 \times 1$ pix$^2$ through $50 \times 50$
pix$^2$, in steps of $1$ pix.  At each step, we identified binned pixels of (1)
high statistical significance ($> 5 \sigma$) in the narrow-band image and (2)
large rest-frame equivalent width ($W_{\rm rest} > 5$ nm for \lya) in the
combination of narrow-band and luminance images, cumulating the identified
pixels at every step.  The sensitivities of the narrow-band, luminance, and
difference images are roughly constant over the central portions of the images
but decrease toward the edges, due to vignetting and dithering.  For
simplicity, we applied the analysis only over the central portions of the
image, specifically over pixel indices ranging in the horizontal direction from
2300 to 9400 and in the vertical direction from 1700 to 7000, where the indices
are zero indexed at the lower-left corner of the image.  The result is a
``detection map'' of 112 features of statistically-significant, strong line
emission that can be reasonably interpreted only as \lya\ of redshift $z
\approx 2.4754$.  The resulting catalog of features is presented in Extended
Data Table 1, which for each feature lists ID number, J2000 Right Ascension and
Declination, energy flux $f$, uncertainty in energy flux $\sigma_f$, radius $r$
(estimated simply from the number of pixels occupied by the feature in the
detection map), and average surface brightness $\langle x \rangle$.  (Note that
the energy flux of each feature includes only whatever portion of the feature
is contained within the detection map.)

\subsection*{\lya\ Surface Brightness Distribution}

Motivated by the intensity distribution of the rest-frame ultraviolet continuum
of high-redshift galaxies\cite{lan2002} and the H$^0$ column density
distribution of QSO absorption lines\cite{pet1993}, we seek to describe the
\lya\ surface brightness distribution $h(x)$ as a function of
surface brightness intensity $x$ by a simple functional form based on a power
law $h(x) \propto x^{-\alpha}$.  But a power law alone cannot suffice, because
for such a distribution, the zeroth moment diverges for $\alpha > 1$ at the
lower limit and the first moment diverges for $\alpha < 2$ at the upper limit.
The divergence of the first moment can be handled by truncating $h(x)$ at
sufficiently large $x$, say by introducing an exponential cutoff
$\exp(-x/x_0)$, where $x_0$ is a characteristic intensity.  The divergence of
the zeroth moment can in principle be handled by in some way truncating $h(x)$
at sufficiently small $x$.  The difficulty is that the normalization of the
zeroth moment is very sensitive to the details of this truncation, which are
hard to measure because small values of $x$ are buried in the noise.

To address this difficulty, we truncate $h(x)$ below some value $x_{\rm min}$,
which is chosen in such a way that it is of order or smaller than the
pixel-to-pixel measurement uncertainty.  We then conceptually consider the
intensities of points on the sky with $x < x_{\rm min}$ to be equivalent to $x
= 0$.  As long as the power-law functional form approximately holds to values
of $x$ as small as (or smaller than) $x_{\rm min}$, then the results are
insensitive to the value of $x_{\rm min}$.  Note that this procedure does not
presume that the distribution is {\em actually} truncated at $x_{\rm min}$, but
rather only that any value $x < x_{\rm min}$ is roughly equivalent to $x = 0$,
given the measurement uncertainty.

Accordingly, we adopt the functional form of the \lya\ surface brightness
distribution $h(x)$ as a function of \lya\ surface brightness intensity $x$ of
equation (1).  Here the first term accounts for all $x < x_{\rm min}$ and
the second term accounts for all $x > x_{\rm min}$ and the ratio $A/B$ is
related to the fraction of the sky covered by emission of $x > x_{\rm min}$.
This procedure serves to ``regularize'' a simple functional form that is
otherwise practically difficult to work with.

We experimented with fitting the observed distribution using a
maximum-likelihood method applied to individual pixels and a $\chi^2$ method
applied to bins of pixels.  We found that both methods yielded essentially
identical results.  For the final analysis, we adopted the $\chi^2$ method,
both because it is computationally far less burdensome than the
maximum-likelihood method and because it provides a simple goodness-of-fit
test.

Due to the drizzling process used to coadd the individual exposures, pixels of
the narrow-band, luminance, and difference images are correlated with their
neighbors.  For the masked difference image, we measured that, as a result of
this correlation, the pixel-to-pixel standard deviation is reduced by a factor
$\approx 0.88$ with respect to the value determined from the propagated
uncertainty image.  This means that the pixels that fall within some $\chi^2$
bin are not statistically independent, which increases the statistical variance
of the bin.  We accounted for this effect using the measured value of the
factor.  As above, we applied the analysis only over the central portions of
the image, i.e.\ over pixel indices ranging in the horizontal direction from
2300 to 9400 and in the vertical direction from 1700 to 7000.

We estimated parameter uncertainties using a bootstrap resampling technique.
Examples of the results are illustrated in Extended Data Fig.\ 5, which shows
the distributions of parameter values obtained in 1000 resamplings for the
parameters $\bar{x}$, $\alpha$, and $x_0$ of the model of equation (2).
(Similar distributions were obtained for the other parameters.)  The
distributions of these parameter values are clearly bimodal, with the values
given in the caption of Fig.\ 2 drawn from the right-most peak in all cases.
We took the parameter uncertainties to be the approximate widths of the
right-most peaks, although we cannot exclude an alternate solution in the other
peaks.  We estimated the uncertainty of the luminosity density using a similar
bootstrap resampling technique.  Higher-sensitivity observations will be needed
to unambiguously measure parameter values.

\subsection*{Image Stacking}

Interpreting a ``stack'' of cutouts of images of many galaxies of similar
redshift involves accounting for a complex blend of the effects of continuum
and line intensity and gas and dust optical depth radial profiles of the target
galaxies and of intensities of background and foreground galaxies.  Here we
explore the interpretation of stacks of broad- and narrow-band galaxy images.

First, consider stacking of broad-band images.  We write the mean intensity
$I(R, z)$ of an ensemble of stacked broad-band galaxy images as a function of
radius $R$ and redshift $z$ as
\begin{equation}
I(R, z) = G_C(R) + I_B(z) \exp{\left[ - \tau_D(R) \right]} + I_F(z),
\end{equation}
where $G_C(R)$ is the target galaxy continuum intensity radial profile
(contributed by starlight, including the possibility of a faint, extended
stellar halo), $I_B(z)$ and $I_F(z)$ are, respectively, the mean background and
foreground intensities (contributed by randomly distributed background and
foreground galaxies), and $\tau_D(R)$ is the target galaxy dust optical depth
radial profile.  In the optically-thin regime appropriate for the outskirts of
galaxies, 
\begin{equation}
I(R, z) \approx G_C(R) + I_B(z) \left[ 1 - \tau_D(R) \right] + I_F(z) \\
= G_C(R) + I_0(z) - I_B(z) \tau_D(R),
\end{equation}
where
\begin{equation}
I_0(z) = I_B(z) + I_F(z).
\end{equation}
In the limit $R \rightarrow 0$, $G(R) \gg I_0(z)$, and
\begin{equation}
I(R, z) \approx G(R).
\end{equation}
In the limit $R \rightarrow \infty$, $G(R) \rightarrow 0$ and $\tau_D(R)
\rightarrow 0$, and
\begin{equation}
I(R, z) \approx I_B(z).
\end{equation}
Equation (4) may be rewritten as
\begin{equation}
\frac{I(R, z)}{I_0(z)} \approx 1 - \frac{I_B(z)}{I_0(z)} \tau_D(R) + \frac{G_C(R)}{I_0(z)}.
\end{equation}
The radial profiles of Fig.\ 4 show $I(R, z)/I_0(z)$ and so may be interpreted
directly in the context of equation (8).  The radial profiles of Fig.\ 4
indicate that $I(R, z)/I_0(z) < 1$ at $R \gtrsim 30$ kpc, from which we
conclude that
\begin{equation}
I_B(z) \tau_D(R) > G_C(R)
\end{equation}
at $z \approx 2.48$ and $R \gtrsim 30$ kpc.  If it is assumed that $I_B(z)
\tau_D(R) \gg G(R)$, then equation (8) provides a way to determine $\tau_D(R)$,
if $I_B(z)$ and $I_F(z)$ are known (through measurement or modeling).  But in
general, even if $I_B(z)$ and $I_F(z)$ are known, then equation (8) indicates
that it will be difficult to untangle $G_C(R)$ (of faint, extended stellar
halos) from $\tau_D(R)$.

Next, consider stacking of narrow-band images.  We write the mean intensity
$I(R, z)$ of an ensemble of stacked narrow-band galaxy images as a function of
radius $R$ and redshift $z$ as
\begin{equation}
I(R, z) = G_C(R) + \Theta(z) G_L(R) + I_B(z) \exp{ \left\{ -\left[ \tau_D(R) +
\Theta(z) \tau_G(R) \right] \right\} } + I_F(z),
\end{equation}
where $G_L(R)$ is the target galaxy line intensity radial profile (contributed
by fluorescence emission), $\tau_G(R)$ is the target galaxy gas optical depth
radial profile, and $\Theta(z)$ is a function that takes on the value
$\Theta(z) = 1$ if the target redshift is an ``on-band'' redshift that shifts
a resonance transition into the filter bandpass and $\Theta(z) = 0$ if the
target redshift is an ``off-band''redshift that does not.  In the
optically-thin regime appropriate for the outskirts of galaxies,
\begin{equation}
I(R, z) \approx G_C(R) + \Theta(z) G_L(R) + I_B(z) \left[ 1 - \tau_D(R) -
\Theta(z) \tau_G(R) \right] + I_F(z),
\end{equation}
Subtracting an adjacent off-band image from an on-band image and assuming that
$G(R)$, $I_B(z)$, and $I_F(z)$ vary slowly with wavelength leaves
\begin{equation}
I(R, z_{\rm on}) - I(R, z_{\rm off}) \approx G_L(R) - I_B(z) \tau_G(R),
\end{equation}
The images and radial profiles of Fig.\ 5 show $I(R, z_{\rm on}) - I(R, z_{\rm
off})$ and so may be interpreted directly in the context of equation (12).  The
radial profiles of Fig.\ 5 indicate that $I(R, z_{\rm on}) - I(R, z_{\rm off})
> 0$ at $R \lesssim 300$ kpc, from which we conclude
\begin{equation}
G_L(R) > I_B(z) \tau_G(R)
\end{equation}
at $z = 2.48$, $1.73$, and $0.51$ for \lya, C~IV, and Mg~II, respectively, and
$R \lesssim 300$ kpc.  If it is assumed that $G_L(R) \gg I_B(z) \tau_G(R)$,
then equation (12) provides a direct measure of $G_L(R)$.  But in general, the
values of the radial profiles shown in Fig.\ 5 represent {\em lower limits} to
$G_L(R)$.

We formed the stacked cutouts of broad- and narrow-band images using a list of
redshifts of galaxies in and surrounding the COSMOS field derived from the
SIMBAD4 v1.8 catalog\cite{wen2000}.

\backmatter

\bmhead{Acknowledgments}

This material is based upon work supported by the National Science Foundation
under Grants 1910001, 2107954, 2108234, 2407763, and 2407764.  We gratefully
acknowledge the staff of Dark Sky New Mexico, including Michael Hensley and
Diana Hensley for their outstanding logistical and technical support and Yuri
Petrunin for crafting six superb instruments.  This research has made use of
the SIMBAD database, operated at CDS, Strasbourg, France.

\bmhead{Data Availability}

All raw Condor data are available following an 18-month proprietary period.
All raw and processed data described here, including the coadded narrow-band,
luminance, and difference images, are available on the Condor web site
https://condorarraytelescope.org/data\_access/ or by contacting the
corresponding author.

\bmhead{Competing Interests}

The authors declare no competing interests.

\bibliography{manuscript}% common bib file


\begin{thebibliography}{32}
\ifx \bisbn   \undefined \def \bisbn  #1{ISBN #1}\fi
\ifx \binits  \undefined \def \binits#1{#1}\fi
\ifx \bauthor  \undefined \def \bauthor#1{#1}\fi
\ifx \batitle  \undefined \def \batitle#1{#1}\fi
\ifx \bjtitle  \undefined \def \bjtitle#1{#1}\fi
\ifx \bvolume  \undefined \def \bvolume#1{\textbf{#1}}\fi
\ifx \byear  \undefined \def \byear#1{#1}\fi
\ifx \bissue  \undefined \def \bissue#1{#1}\fi
\ifx \bfpage  \undefined \def \bfpage#1{#1}\fi
\ifx \blpage  \undefined \def \blpage #1{#1}\fi
\ifx \burl  \undefined \def \burl#1{\textsf{#1}}\fi
\ifx \doiurl  \undefined \def \doiurl#1{\url{https://doi.org/#1}}\fi
\ifx \betal  \undefined \def \betal{\textit{et al.}}\fi
\ifx \binstitute  \undefined \def \binstitute#1{#1}\fi
\ifx \binstitutionaled  \undefined \def \binstitutionaled#1{#1}\fi
\ifx \bctitle  \undefined \def \bctitle#1{#1}\fi
\ifx \beditor  \undefined \def \beditor#1{#1}\fi
\ifx \bpublisher  \undefined \def \bpublisher#1{#1}\fi
\ifx \bbtitle  \undefined \def \bbtitle#1{#1}\fi
\ifx \bedition  \undefined \def \bedition#1{#1}\fi
\ifx \bseriesno  \undefined \def \bseriesno#1{#1}\fi
\ifx \blocation  \undefined \def \blocation#1{#1}\fi
\ifx \bsertitle  \undefined \def \bsertitle#1{#1}\fi
\ifx \bsnm \undefined \def \bsnm#1{#1}\fi
\ifx \bsuffix \undefined \def \bsuffix#1{#1}\fi
\ifx \bparticle \undefined \def \bparticle#1{#1}\fi
\ifx \barticle \undefined \def \barticle#1{#1}\fi
\bibcommenthead
\ifx \bconfdate \undefined \def \bconfdate #1{#1}\fi
\ifx \botherref \undefined \def \botherref #1{#1}\fi
\ifx \url \undefined \def \url#1{\textsf{#1}}\fi
\ifx \bchapter \undefined \def \bchapter#1{#1}\fi
\ifx \bbook \undefined \def \bbook#1{#1}\fi
\ifx \bcomment \undefined \def \bcomment#1{#1}\fi
\ifx \oauthor \undefined \def \oauthor#1{#1}\fi
\ifx \citeauthoryear \undefined \def \citeauthoryear#1{#1}\fi
\ifx \endbibitem  \undefined \def \endbibitem {}\fi
\ifx \bconflocation  \undefined \def \bconflocation#1{#1}\fi
\ifx \arxivurl  \undefined \def \arxivurl#1{\textsf{#1}}\fi
\csname PreBibitemsHook\endcsname

%%% 1
\bibitem[\protect\citeauthoryear{{Cen} and {Ostriker}}{2006}]{cen2006}
\begin{barticle}
\bauthor{\bsnm{{Cen}}, \binits{R.}},
\bauthor{\bsnm{{Ostriker}}, \binits{J.P.}}:
\batitle{{Where Are the Baryons? II. Feedback Effects}}.
\bjtitle{\apj}
\bvolume{650}(\bissue{2}),
\bfpage{560}--\blpage{572}
(\byear{2006})
\doiurl{10.1086/506505}
{\href{https://arxiv.org/abs/astro-ph/0601008}{{arXiv:astro-ph/0601008}}}
{[astro-ph]}
\end{barticle}
\endbibitem

%%% 2
\bibitem[\protect\citeauthoryear{{Tumlinson} et~al.}{2017}]{tum2017}
\begin{barticle}
\bauthor{\bsnm{{Tumlinson}}, \binits{J.}},
\bauthor{\bsnm{{Peeples}}, \binits{M.S.}},
\bauthor{\bsnm{{Werk}}, \binits{J.K.}}:
\batitle{{The Circumgalactic Medium}}.
\bjtitle{\araa}
\bvolume{55}(\bissue{1}),
\bfpage{389}--\blpage{432}
(\byear{2017})
\doiurl{10.1146/annurev-astro-091916-055240}
{\href{https://arxiv.org/abs/1709.09180}{{arXiv:1709.09180}}}
{[astro-ph.GA]}
\end{barticle}
\endbibitem

%%% 3
\bibitem[\protect\citeauthoryear{{Cantalupo} et~al.}{2014}]{can2014}
\begin{barticle}
\bauthor{\bsnm{{Cantalupo}}, \binits{S.}},
\bauthor{\bsnm{{Arrigoni-Battaia}}, \binits{F.}},
\bauthor{\bsnm{{Prochaska}}, \binits{J.X.}},
\bauthor{\bsnm{{Hennawi}}, \binits{J.F.}},
\bauthor{\bsnm{{Madau}}, \binits{P.}}:
\batitle{{A cosmic web filament revealed in Lyman-{\ensuremath{\alpha}}
  emission around a luminous high-redshift quasar}}.
\bjtitle{\nat}
\bvolume{506}(\bissue{7486}),
\bfpage{63}--\blpage{66}
(\byear{2014})
\doiurl{10.1038/nature12898}
{\href{https://arxiv.org/abs/1401.4469}{{arXiv:1401.4469}}}
{[astro-ph.CO]}
\end{barticle}
\endbibitem

%%% 4
\bibitem[\protect\citeauthoryear{{Gallego} et~al.}{2018}]{gal2018}
\begin{barticle}
\bauthor{\bsnm{{Gallego}}, \binits{S.G.}},
\bauthor{\bsnm{{Cantalupo}}, \binits{S.}},
\bauthor{\bsnm{{Lilly}}, \binits{S.}},
\bauthor{\bsnm{{Marino}}, \binits{R.A.}},
\bauthor{\bsnm{{Pezzulli}}, \binits{G.}},
\bauthor{\bsnm{{Schaye}}, \binits{J.}},
\bauthor{\bsnm{{Wisotzki}}, \binits{L.}},
\bauthor{\bsnm{{Bacon}}, \binits{R.}},
\bauthor{\bsnm{{Inami}}, \binits{H.}},
\bauthor{\bsnm{{Akhlaghi}}, \binits{M.}},
\bauthor{\bsnm{{Tacchella}}, \binits{S.}},
\bauthor{\bsnm{{Richard}}, \binits{J.}},
\bauthor{\bsnm{{Bouche}}, \binits{N.F.}},
\bauthor{\bsnm{{Steinmetz}}, \binits{M.}},
\bauthor{\bsnm{{Carollo}}, \binits{M.}}:
\batitle{{Stacking the Cosmic Web in fluorescent Ly {\ensuremath{\alpha}}
  emission with MUSE}}.
\bjtitle{\mnras}
\bvolume{475}(\bissue{3}),
\bfpage{3854}--\blpage{3869}
(\byear{2018})
\doiurl{10.1093/mnras/sty037}
{\href{https://arxiv.org/abs/1706.03785}{{arXiv:1706.03785}}}
{[astro-ph.CO]}
\end{barticle}
\endbibitem

%%% 5
\bibitem[\protect\citeauthoryear{{Bacon} et~al.}{2021}]{bac2021}
\begin{barticle}
\bauthor{\bsnm{{Bacon}}, \binits{R.}},
\bauthor{\bsnm{{Mary}}, \binits{D.}},
\bauthor{\bsnm{{Garel}}, \binits{T.}},
\bauthor{\bsnm{{Blaizot}}, \binits{J.}},
\bauthor{\bsnm{{Maseda}}, \binits{M.}},
\bauthor{\bsnm{{Schaye}}, \binits{J.}},
\bauthor{\bsnm{{Wisotzki}}, \binits{L.}},
\bauthor{\bsnm{{Conseil}}, \binits{S.}},
\bauthor{\bsnm{{Brinchmann}}, \binits{J.}},
\bauthor{\bsnm{{Leclercq}}, \binits{F.}},
\bauthor{\bsnm{{Abril-Melgarejo}}, \binits{V.}},
\bauthor{\bsnm{{Boogaard}}, \binits{L.}},
\bauthor{\bsnm{{Bouch{\'e}}}, \binits{N.F.}},
\bauthor{\bsnm{{Contini}}, \binits{T.}},
\bauthor{\bsnm{{Feltre}}, \binits{A.}},
\bauthor{\bsnm{{Guiderdoni}}, \binits{B.}},
\bauthor{\bsnm{{Herenz}}, \binits{C.}},
\bauthor{\bsnm{{Kollatschny}}, \binits{W.}},
\bauthor{\bsnm{{Kusakabe}}, \binits{H.}},
\bauthor{\bsnm{{Matthee}}, \binits{J.}},
\bauthor{\bsnm{{Michel-Dansac}}, \binits{L.}},
\bauthor{\bsnm{{Nanayakkara}}, \binits{T.}},
\bauthor{\bsnm{{Richard}}, \binits{J.}},
\bauthor{\bsnm{{Roth}}, \binits{M.}},
\bauthor{\bsnm{{Schmidt}}, \binits{K.B.}},
\bauthor{\bsnm{{Steinmetz}}, \binits{M.}},
\bauthor{\bsnm{{Tresse}}, \binits{L.}},
\bauthor{\bsnm{{Urrutia}}, \binits{T.}},
\bauthor{\bsnm{{Verhamme}}, \binits{A.}},
\bauthor{\bsnm{{Weilbacher}}, \binits{P.M.}},
\bauthor{\bsnm{{Zabl}}, \binits{J.}},
\bauthor{\bsnm{{Zoutendijk}}, \binits{S.L.}}:
\batitle{{The MUSE Extremely Deep Field: The cosmic web in emission at high
  redshift}}.
\bjtitle{\aap}
\bvolume{647},
\bfpage{107}
(\byear{2021})
\doiurl{10.1051/0004-6361/202039887}
{\href{https://arxiv.org/abs/2102.05516}{{arXiv:2102.05516}}}
{[astro-ph.CO]}
\end{barticle}
\endbibitem

%%% 6
\bibitem[\protect\citeauthoryear{Martin et~al.}{2023}]{mar2023}
\begin{botherref}
\oauthor{\bsnm{Martin}, \binits{C.}},
\oauthor{\bsnm{Darvish}, \binits{B.}},
\oauthor{\bsnm{Lin}, \binits{Z.}},
\oauthor{\bsnm{Cen}, \binits{R.}},
\oauthor{\bsnm{Matuszewski}, \binits{M.}},
\oauthor{\bsnm{Morrissey}, \binits{P.}},
\oauthor{\bsnm{Neill}, \binits{J.}},
\oauthor{\bsnm{Moore}, \binits{A.}}:
Extensive diffuse lyman-alpha emission correlated with cosmic structure.
Nature Astronomy,
1--12
(2023)
\doiurl{10.1038/s41550-023-02054-1}
\end{botherref}
\endbibitem

%%% 7
\bibitem[\protect\citeauthoryear{{Lanzetta} et~al.}{2023}]{lan2023}
\begin{barticle}
\bauthor{\bsnm{{Lanzetta}}, \binits{K.M.}},
\bauthor{\bsnm{{Gromoll}}, \binits{S.}},
\bauthor{\bsnm{{Shara}}, \binits{M.M.}},
\bauthor{\bsnm{{Berg}}, \binits{S.}},
\bauthor{\bsnm{{Valls-Gabaud}}, \binits{D.}},
\bauthor{\bsnm{{Walter}}, \binits{F.M.}},
\bauthor{\bsnm{{Webb}}, \binits{J.K.}}:
\batitle{{Introducing the Condor Array Telescope. I. Motivation, Configuration,
  and Performance}}.
\bjtitle{\pasp}
\bvolume{135}(\bissue{1043}),
\bfpage{015002}
(\byear{2023})
\doiurl{10.1088/1538-3873/acaee6}
{\href{https://arxiv.org/abs/2301.06301}{{arXiv:2301.06301}}}
{[astro-ph.IM]}
\end{barticle}
\endbibitem

%%% 8
\bibitem[\protect\citeauthoryear{{Hazard} et~al.}{1963}]{haz1963}
\begin{barticle}
\bauthor{\bsnm{{Hazard}}, \binits{C.}},
\bauthor{\bsnm{{Mackey}}, \binits{M.B.}},
\bauthor{\bsnm{{Shimmins}}, \binits{A.J.}}:
\batitle{{Investigation of the Radio Source 3C 273 By The Method of Lunar
  Occultations}}.
\bjtitle{\nat}
\bvolume{197}(\bissue{4872}),
\bfpage{1037}--\blpage{1039}
(\byear{1963})
\doiurl{10.1038/1971037a0}
\end{barticle}
\endbibitem

%%% 9
\bibitem[\protect\citeauthoryear{{Schmidt}}{1963}]{sch1963}
\begin{barticle}
\bauthor{\bsnm{{Schmidt}}, \binits{M.}}:
\batitle{{3C 273 : A Star-Like Object with Large Red-Shift}}.
\bjtitle{\nat}
\bvolume{197}(\bissue{4872}),
\bfpage{1040}
(\byear{1963})
\doiurl{10.1038/1971040a0}
\end{barticle}
\endbibitem

%%% 10
\bibitem[\protect\citeauthoryear{{Lynds}}{1971}]{lyn1971}
\begin{barticle}
\bauthor{\bsnm{{Lynds}}, \binits{R.}}:
\batitle{{The Absorption-Line Spectrum of 4c 05.34}}.
\bjtitle{ApJL}
\bvolume{164},
\bfpage{73}
(\byear{1971})
\doiurl{10.1086/180695}
\end{barticle}
\endbibitem

%%% 11
\bibitem[\protect\citeauthoryear{{Bahcall} and {Spitzer}}{1969}]{bah1969}
\begin{barticle}
\bauthor{\bsnm{{Bahcall}}, \binits{J.N.}},
\bauthor{\bsnm{{Spitzer}}, \binits{J.} \bsuffix{Lyman}}:
\batitle{{Absorption Lines Produced by Galactic Halos}}.
\bjtitle{\apjl}
\bvolume{156},
\bfpage{63}
(\byear{1969})
\doiurl{10.1086/180350}
\end{barticle}
\endbibitem

%%% 12
\bibitem[\protect\citeauthoryear{{Sargent} et~al.}{1980}]{sar1980}
\begin{barticle}
\bauthor{\bsnm{{Sargent}}, \binits{W.L.W.}},
\bauthor{\bsnm{{Young}}, \binits{P.J.}},
\bauthor{\bsnm{{Boksenberg}}, \binits{A.}},
\bauthor{\bsnm{{Tytler}}, \binits{D.}}:
\batitle{{The distribution of Lyman-alpha absorption lines in the spectra of
  six QSOs: evidence for an intergalactic origin.}}
\bjtitle{\apjs}
\bvolume{42},
\bfpage{41}--\blpage{81}
(\byear{1980})
\doiurl{10.1086/190644}
\end{barticle}
\endbibitem

%%% 13
\bibitem[\protect\citeauthoryear{{Schaye} et~al.}{2015}]{sch2015}
\begin{barticle}
\bauthor{\bsnm{{Schaye}}, \binits{J.}},
\bauthor{\bsnm{{Crain}}, \binits{R.A.}},
\bauthor{\bsnm{{Bower}}, \binits{R.G.}},
\bauthor{\bsnm{{Furlong}}, \binits{M.}},
\bauthor{\bsnm{{Schaller}}, \binits{M.}},
\bauthor{\bsnm{{Theuns}}, \binits{T.}},
\bauthor{\bsnm{{Dalla Vecchia}}, \binits{C.}},
\bauthor{\bsnm{{Frenk}}, \binits{C.S.}},
\bauthor{\bsnm{{McCarthy}}, \binits{I.G.}},
\bauthor{\bsnm{{Helly}}, \binits{J.C.}},
\bauthor{\bsnm{{Jenkins}}, \binits{A.}},
\bauthor{\bsnm{{Rosas-Guevara}}, \binits{Y.M.}},
\bauthor{\bsnm{{White}}, \binits{S.D.M.}},
\bauthor{\bsnm{{Baes}}, \binits{M.}},
\bauthor{\bsnm{{Booth}}, \binits{C.M.}},
\bauthor{\bsnm{{Camps}}, \binits{P.}},
\bauthor{\bsnm{{Navarro}}, \binits{J.F.}},
\bauthor{\bsnm{{Qu}}, \binits{Y.}},
\bauthor{\bsnm{{Rahmati}}, \binits{A.}},
\bauthor{\bsnm{{Sawala}}, \binits{T.}},
\bauthor{\bsnm{{Thomas}}, \binits{P.A.}},
\bauthor{\bsnm{{Trayford}}, \binits{J.}}:
\batitle{{The EAGLE project: simulating the evolution and assembly of galaxies
  and their environments}}.
\bjtitle{MNRAS}
\bvolume{446}(\bissue{1}),
\bfpage{521}--\blpage{554}
(\byear{2015})
\doiurl{10.1093/mnras/stu2058}
{\href{https://arxiv.org/abs/1407.7040}{{arXiv:1407.7040}}}
{[astro-ph.GA]}
\end{barticle}
\endbibitem

%%% 14
\bibitem[\protect\citeauthoryear{{Bolton} et~al.}{2017}]{bol2017}
\begin{barticle}
\bauthor{\bsnm{{Bolton}}, \binits{J.S.}},
\bauthor{\bsnm{{Puchwein}}, \binits{E.}},
\bauthor{\bsnm{{Sijacki}}, \binits{D.}},
\bauthor{\bsnm{{Haehnelt}}, \binits{M.G.}},
\bauthor{\bsnm{{Kim}}, \binits{T.-S.}},
\bauthor{\bsnm{{Meiksin}}, \binits{A.}},
\bauthor{\bsnm{{Regan}}, \binits{J.A.}},
\bauthor{\bsnm{{Viel}}, \binits{M.}}:
\batitle{{The Sherwood simulation suite: overview and data comparisons with the
  Lyman {\ensuremath{\alpha}} forest at redshifts 2 {\ensuremath{\leq}} z
  {\ensuremath{\leq}} 5}}.
\bjtitle{\mnras}
\bvolume{464}(\bissue{1}),
\bfpage{897}--\blpage{914}
(\byear{2017})
\doiurl{10.1093/mnras/stw2397}
{\href{https://arxiv.org/abs/1605.03462}{{arXiv:1605.03462}}}
{[astro-ph.CO]}
\end{barticle}
\endbibitem

%%% 15
\bibitem[\protect\citeauthoryear{{Hogan} and {Weymann}}{1987}]{hog1987}
\begin{barticle}
\bauthor{\bsnm{{Hogan}}, \binits{C.J.}},
\bauthor{\bsnm{{Weymann}}, \binits{R.J.}}:
\batitle{{Lyman-alpha emission from the Lyman-alpha forest}}.
\bjtitle{MNRAS}
\bvolume{225},
\bfpage{1}--\blpage{5}
(\byear{1987})
\doiurl{10.1093/mnras/225.1.1P}
\end{barticle}
\endbibitem

%%% 16
\bibitem[\protect\citeauthoryear{{Cantalupo} et~al.}{2005}]{can2005}
\begin{barticle}
\bauthor{\bsnm{{Cantalupo}}, \binits{S.}},
\bauthor{\bsnm{{Porciani}}, \binits{C.}},
\bauthor{\bsnm{{Lilly}}, \binits{S.J.}},
\bauthor{\bsnm{{Miniati}}, \binits{F.}}:
\batitle{{Fluorescent Ly{\ensuremath{\alpha}} Emission from the High-Redshift
  Intergalactic Medium}}.
\bjtitle{\apj}
\bvolume{628}(\bissue{1}),
\bfpage{61}--\blpage{75}
(\byear{2005})
\doiurl{10.1086/430758}
{\href{https://arxiv.org/abs/astro-ph/0504015}{{arXiv:astro-ph/0504015}}}
{[astro-ph]}
\end{barticle}
\endbibitem

%%% 17
\bibitem[\protect\citeauthoryear{{Scoville} et~al.}{2013}]{sco2013}
\begin{barticle}
\bauthor{\bsnm{{Scoville}}, \binits{N.}},
\bauthor{\bsnm{{Arnouts}}, \binits{S.}},
\bauthor{\bsnm{{Aussel}}, \binits{H.}},
\bauthor{\bsnm{{Benson}}, \binits{A.}},
\bauthor{\bsnm{{Bongiorno}}, \binits{A.}},
\bauthor{\bsnm{{Bundy}}, \binits{K.}},
\bauthor{\bsnm{{Calvo}}, \binits{M.A.A.}},
\bauthor{\bsnm{{Capak}}, \binits{P.}},
\bauthor{\bsnm{{Carollo}}, \binits{M.}},
\bauthor{\bsnm{{Civano}}, \binits{F.}},
\bauthor{\bsnm{{Dunlop}}, \binits{J.}},
\bauthor{\bsnm{{Elvis}}, \binits{M.}},
\bauthor{\bsnm{{Faisst}}, \binits{A.}},
\bauthor{\bsnm{{Finoguenov}}, \binits{A.}},
\bauthor{\bsnm{{Fu}}, \binits{H.}},
\bauthor{\bsnm{{Giavalisco}}, \binits{M.}},
\bauthor{\bsnm{{Guo}}, \binits{Q.}},
\bauthor{\bsnm{{Ilbert}}, \binits{O.}},
\bauthor{\bsnm{{Iovino}}, \binits{A.}},
\bauthor{\bsnm{{Kajisawa}}, \binits{M.}},
\bauthor{\bsnm{{Kartaltepe}}, \binits{J.}},
\bauthor{\bsnm{{Leauthaud}}, \binits{A.}},
\bauthor{\bsnm{{Le F{\`e}vre}}, \binits{O.}},
\bauthor{\bsnm{{LeFloch}}, \binits{E.}},
\bauthor{\bsnm{{Lilly}}, \binits{S.J.}},
\bauthor{\bsnm{{Liu}}, \binits{C.T.-C.}},
\bauthor{\bsnm{{Manohar}}, \binits{S.}},
\bauthor{\bsnm{{Massey}}, \binits{R.}},
\bauthor{\bsnm{{Masters}}, \binits{D.}},
\bauthor{\bsnm{{McCracken}}, \binits{H.J.}},
\bauthor{\bsnm{{Mobasher}}, \binits{B.}},
\bauthor{\bsnm{{Peng}}, \binits{Y.-J.}},
\bauthor{\bsnm{{Renzini}}, \binits{A.}},
\bauthor{\bsnm{{Rhodes}}, \binits{J.}},
\bauthor{\bsnm{{Salvato}}, \binits{M.}},
\bauthor{\bsnm{{Sanders}}, \binits{D.B.}},
\bauthor{\bsnm{{Sarvestani}}, \binits{B.D.}},
\bauthor{\bsnm{{Scarlata}}, \binits{C.}},
\bauthor{\bsnm{{Schinnerer}}, \binits{E.}},
\bauthor{\bsnm{{Sheth}}, \binits{K.}},
\bauthor{\bsnm{{Shopbell}}, \binits{P.L.}},
\bauthor{\bsnm{{Smol{\v{c}}i{\'c}}}, \binits{V.}},
\bauthor{\bsnm{{Taniguchi}}, \binits{Y.}},
\bauthor{\bsnm{{Taylor}}, \binits{J.E.}},
\bauthor{\bsnm{{White}}, \binits{S.D.M.}},
\bauthor{\bsnm{{Yan}}, \binits{L.}}:
\batitle{{Evolution of Galaxies and Their Environments at z = 0.1-3 in
  COSMOS}}.
\bjtitle{\apjs}
\bvolume{206}(\bissue{1}),
\bfpage{3}
(\byear{2013})
\doiurl{10.1088/0067-0049/206/1/3}
{\href{https://arxiv.org/abs/1303.6689}{{arXiv:1303.6689}}}
{[astro-ph.CO]}
\end{barticle}
\endbibitem

%%% 18
\bibitem[\protect\citeauthoryear{{Cucciati} et~al.}{2018}]{cuc2018}
\begin{barticle}
\bauthor{\bsnm{{Cucciati}}, \binits{O.}},
\bauthor{\bsnm{{Lemaux}}, \binits{B.C.}},
\bauthor{\bsnm{{Zamorani}}, \binits{G.}},
\bauthor{\bsnm{{Le F{\`e}vre}}, \binits{O.}},
\bauthor{\bsnm{{Tasca}}, \binits{L.A.M.}},
\bauthor{\bsnm{{Hathi}}, \binits{N.P.}},
\bauthor{\bsnm{{Lee}}, \binits{K.-G.}},
\bauthor{\bsnm{{Bardelli}}, \binits{S.}},
\bauthor{\bsnm{{Cassata}}, \binits{P.}},
\bauthor{\bsnm{{Garilli}}, \binits{B.}},
\bauthor{\bsnm{{Le Brun}}, \binits{V.}},
\bauthor{\bsnm{{Maccagni}}, \binits{D.}},
\bauthor{\bsnm{{Pentericci}}, \binits{L.}},
\bauthor{\bsnm{{Thomas}}, \binits{R.}},
\bauthor{\bsnm{{Vanzella}}, \binits{E.}},
\bauthor{\bsnm{{Zucca}}, \binits{E.}},
\bauthor{\bsnm{{Lubin}}, \binits{L.M.}},
\bauthor{\bsnm{{Amorin}}, \binits{R.}},
\bauthor{\bsnm{{Cassar{\`a}}}, \binits{L.P.}},
\bauthor{\bsnm{{Cimatti}}, \binits{A.}},
\bauthor{\bsnm{{Talia}}, \binits{M.}},
\bauthor{\bsnm{{Vergani}}, \binits{D.}},
\bauthor{\bsnm{{Koekemoer}}, \binits{A.}},
\bauthor{\bsnm{{Pforr}}, \binits{J.}},
\bauthor{\bsnm{{Salvato}}, \binits{M.}}:
\batitle{{The progeny of a cosmic titan: a massive multi-component
  proto-supercluster in formation at $z = 2.45$ in VUDS}}.
\bjtitle{\aap}
\bvolume{619},
\bfpage{49}
(\byear{2018})
\doiurl{10.1051/0004-6361/201833655}
{\href{https://arxiv.org/abs/1806.06073}{{arXiv:1806.06073}}}
{[astro-ph.CO]}
\end{barticle}
\endbibitem

%%% 19
\bibitem[\protect\citeauthoryear{{Lanzetta} et~al.}{2023}]{lan2023a}
\begin{botherref}
\oauthor{\bsnm{{Lanzetta}}, \binits{K.M.}},
\oauthor{\bsnm{{Gromoll}}, \binits{S.}},
\oauthor{\bsnm{{Shara}}, \binits{M.M.}},
\oauthor{\bsnm{{Berg}}, \binits{S.}},
\oauthor{\bsnm{{Garland}}, \binits{J.}},
\oauthor{\bsnm{{Mancini}}, \binits{E.}},
\oauthor{\bsnm{{Valls-Gabaud}}, \binits{D.}},
\oauthor{\bsnm{{Walter}}, \binits{F.M.}},
\oauthor{\bsnm{{Webb}}, \binits{J.K.}}:
{Introducing the Condor Array Telescope. II. Deep imaging observations of the
  edge-on spiral galaxy NGC 5907 and the NGC 5866 Group: yet another view of
  the iconic stellar stream}.
\mnras
(2023)
\end{botherref}
\endbibitem

%%% 20
\bibitem[\protect\citeauthoryear{Turin}{1960}]{tur1960}
\begin{barticle}
\bauthor{\bsnm{Turin}, \binits{G.}}:
\batitle{An introduction to matched filters}.
\bjtitle{IRE Transactions on Information Theory}
\bvolume{6}(\bissue{3}),
\bfpage{311}--\blpage{329}
(\byear{1960})
\doiurl{10.1109/TIT.1960.1057571}
\end{barticle}
\endbibitem

%%% 21
\bibitem[\protect\citeauthoryear{{Trump} et~al.}{2009}]{tru2009}
\begin{barticle}
\bauthor{\bsnm{{Trump}}, \binits{J.R.}},
\bauthor{\bsnm{{Impey}}, \binits{C.D.}},
\bauthor{\bsnm{{Elvis}}, \binits{M.}},
\bauthor{\bsnm{{McCarthy}}, \binits{P.J.}},
\bauthor{\bsnm{{Huchra}}, \binits{J.P.}},
\bauthor{\bsnm{{Brusa}}, \binits{M.}},
\bauthor{\bsnm{{Salvato}}, \binits{M.}},
\bauthor{\bsnm{{Capak}}, \binits{P.}},
\bauthor{\bsnm{{Cappelluti}}, \binits{N.}},
\bauthor{\bsnm{{Civano}}, \binits{F.}},
\bauthor{\bsnm{{Comastri}}, \binits{A.}},
\bauthor{\bsnm{{Gabor}}, \binits{J.}},
\bauthor{\bsnm{{Hao}}, \binits{H.}},
\bauthor{\bsnm{{Hasinger}}, \binits{G.}},
\bauthor{\bsnm{{Jahnke}}, \binits{K.}},
\bauthor{\bsnm{{Kelly}}, \binits{B.C.}},
\bauthor{\bsnm{{Lilly}}, \binits{S.J.}},
\bauthor{\bsnm{{Schinnerer}}, \binits{E.}},
\bauthor{\bsnm{{Scoville}}, \binits{N.Z.}},
\bauthor{\bsnm{{Smol{\v{c}}i{\'c}}}, \binits{V.}}:
\batitle{{The COSMOS Active Galactic Nucleus Spectroscopic Survey. I.
  XMM-Newton Counterparts}}.
\bjtitle{\apj}
\bvolume{696}(\bissue{2}),
\bfpage{1195}--\blpage{1212}
(\byear{2009})
\doiurl{10.1088/0004-637X/696/2/1195}
{\href{https://arxiv.org/abs/0811.3977}{{arXiv:0811.3977}}}
{[astro-ph]}
\end{barticle}
\endbibitem

%%% 22
\bibitem[\protect\citeauthoryear{{Lanzetta} et~al.}{2002}]{lan2002}
\begin{barticle}
\bauthor{\bsnm{{Lanzetta}}, \binits{K.M.}},
\bauthor{\bsnm{{Yahata}}, \binits{N.}},
\bauthor{\bsnm{{Pascarelle}}, \binits{S.}},
\bauthor{\bsnm{{Chen}}, \binits{H.-W.}},
\bauthor{\bsnm{{Fern{\'a}ndez-Soto}}, \binits{A.}}:
\batitle{{The Star Formation Rate Intensity Distribution Function: Implications
  for the Cosmic Star Formation Rate History of the Universe}}.
\bjtitle{\apj}
\bvolume{570}(\bissue{2}),
\bfpage{492}--\blpage{501}
(\byear{2002})
\doiurl{10.1086/339774}
{\href{https://arxiv.org/abs/astro-ph/0111129}{{arXiv:astro-ph/0111129}}}
{[astro-ph]}
\end{barticle}
\endbibitem

%%% 23
\bibitem[\protect\citeauthoryear{{Petitjean} et~al.}{1993}]{pet1993}
\begin{barticle}
\bauthor{\bsnm{{Petitjean}}, \binits{P.}},
\bauthor{\bsnm{{Webb}}, \binits{J.K.}},
\bauthor{\bsnm{{Rauch}}, \binits{M.}},
\bauthor{\bsnm{{Carswell}}, \binits{R.F.}},
\bauthor{\bsnm{{Lanzetta}}, \binits{K.}}:
\batitle{{Evidence for structure in the H I column density distribution of QSO
  absorbers.}}
\bjtitle{\mnras}
\bvolume{262},
\bfpage{499}--\blpage{505}
(\byear{1993})
\doiurl{10.1093/mnras/262.2.499}
\end{barticle}
\endbibitem

%%% 24
\bibitem[\protect\citeauthoryear{{Taniguchi} et~al.}{2015}]{tan2015}
\begin{barticle}
\bauthor{\bsnm{{Taniguchi}}, \binits{Y.}},
\bauthor{\bsnm{{Kajisawa}}, \binits{M.}},
\bauthor{\bsnm{{Kobayashi}}, \binits{M.A.R.}},
\bauthor{\bsnm{{Shioya}}, \binits{Y.}},
\bauthor{\bsnm{{Nagao}}, \binits{T.}},
\bauthor{\bsnm{{Capak}}, \binits{P.L.}},
\bauthor{\bsnm{{Aussel}}, \binits{H.}},
\bauthor{\bsnm{{Ichikawa}}, \binits{A.}},
\bauthor{\bsnm{{Murayama}}, \binits{T.}},
\bauthor{\bsnm{{Scoville}}, \binits{N.Z.}},
\bauthor{\bsnm{{Ilbert}}, \binits{O.}},
\bauthor{\bsnm{{Salvato}}, \binits{M.}},
\bauthor{\bsnm{{Sanders}}, \binits{D.B.B.}},
\bauthor{\bsnm{{Mobasher}}, \binits{B.}},
\bauthor{\bsnm{{Miyazaki}}, \binits{S.}},
\bauthor{\bsnm{{Komiyama}}, \binits{Y.}},
\bauthor{\bsnm{{Le F{\`e}vre}}, \binits{O.}},
\bauthor{\bsnm{{Tasca}}, \binits{L.}},
\bauthor{\bsnm{{Lilly}}, \binits{S.}},
\bauthor{\bsnm{{Carollo}}, \binits{M.}},
\bauthor{\bsnm{{Renzini}}, \binits{A.}},
\bauthor{\bsnm{{Rich}}, \binits{M.}},
\bauthor{\bsnm{{Schinnerer}}, \binits{E.}},
\bauthor{\bsnm{{Kaifu}}, \binits{N.}},
\bauthor{\bsnm{{Karoji}}, \binits{H.}},
\bauthor{\bsnm{{Arimoto}}, \binits{N.}},
\bauthor{\bsnm{{Okamura}}, \binits{S.}},
\bauthor{\bsnm{{Ohta}}, \binits{K.}},
\bauthor{\bsnm{{Shimasaku}}, \binits{K.}},
\bauthor{\bsnm{{Hayashino}}, \binits{T.}}:
\batitle{{The Subaru COSMOS 20: Subaru optical imaging of the HST COSMOS field
  with 20 filters*}}.
\bjtitle{\pasj}
\bvolume{67}(\bissue{6}),
\bfpage{104}
(\byear{2015})
\doiurl{10.1093/pasj/psv106}
{\href{https://arxiv.org/abs/1510.00550}{{arXiv:1510.00550}}}
{[astro-ph.GA]}
\end{barticle}
\endbibitem

%%% 25
\bibitem[\protect\citeauthoryear{{York} et~al.}{2006}]{yor2006}
\begin{barticle}
\bauthor{\bsnm{{York}}, \binits{D.G.}},
\bauthor{\bsnm{{Khare}}, \binits{P.}},
\bauthor{\bsnm{{Vanden Berk}}, \binits{D.}},
\bauthor{\bsnm{{Kulkarni}}, \binits{V.P.}},
\bauthor{\bsnm{{Crotts}}, \binits{A.P.S.}},
\bauthor{\bsnm{{Lauroesch}}, \binits{J.T.}},
\bauthor{\bsnm{{Richards}}, \binits{G.T.}},
\bauthor{\bsnm{{Schneider}}, \binits{D.P.}},
\bauthor{\bsnm{{Welty}}, \binits{D.E.}},
\bauthor{\bsnm{{Alsayyad}}, \binits{Y.}},
\bauthor{\bsnm{{Kumar}}, \binits{A.}},
\bauthor{\bsnm{{Lundgren}}, \binits{B.}},
\bauthor{\bsnm{{Shanidze}}, \binits{N.}},
\bauthor{\bsnm{{Smith}}, \binits{T.}},
\bauthor{\bsnm{{Vanlandingham}}, \binits{J.}},
\bauthor{\bsnm{{Baugher}}, \binits{B.}},
\bauthor{\bsnm{{Hall}}, \binits{P.B.}},
\bauthor{\bsnm{{Jenkins}}, \binits{E.B.}},
\bauthor{\bsnm{{Menard}}, \binits{B.}},
\bauthor{\bsnm{{Rao}}, \binits{S.}},
\bauthor{\bsnm{{Tumlinson}}, \binits{J.}},
\bauthor{\bsnm{{Turnshek}}, \binits{D.}},
\bauthor{\bsnm{{Yip}}, \binits{C.-W.}},
\bauthor{\bsnm{{Brinkmann}}, \binits{J.}}:
\batitle{{Average extinction curves and relative abundances for quasi-stellar
  object absorption-line systems at $1 \leq z_{\rm abs} < 2$}}.
\bjtitle{\mnras}
\bvolume{367}(\bissue{3}),
\bfpage{945}--\blpage{978}
(\byear{2006})
\doiurl{10.1111/j.1365-2966.2005.10018.x}
{\href{https://arxiv.org/abs/astro-ph/0601279}{{arXiv:astro-ph/0601279}}}
{[astro-ph]}
\end{barticle}
\endbibitem

%%% 26
\bibitem[\protect\citeauthoryear{{M{\'e}nard} et~al.}{2010}]{men2010}
\begin{barticle}
\bauthor{\bsnm{{M{\'e}nard}}, \binits{B.}},
\bauthor{\bsnm{{Scranton}}, \binits{R.}},
\bauthor{\bsnm{{Fukugita}}, \binits{M.}},
\bauthor{\bsnm{{Richards}}, \binits{G.}}:
\batitle{{Measuring the galaxy-mass and galaxy-dust correlations through
  magnification and reddening}}.
\bjtitle{\mnras}
\bvolume{405}(\bissue{2}),
\bfpage{1025}--\blpage{1039}
(\byear{2010})
\doiurl{10.1111/j.1365-2966.2010.16486.x}
{\href{https://arxiv.org/abs/0902.4240}{{arXiv:0902.4240}}}
{[astro-ph.CO]}
\end{barticle}
\endbibitem

%%% 27
\bibitem[\protect\citeauthoryear{{Peebles} and {Groth}}{1975}]{pee1975}
\begin{barticle}
\bauthor{\bsnm{{Peebles}}, \binits{P.J.E.}},
\bauthor{\bsnm{{Groth}}, \binits{E.J.}}:
\batitle{{Statistical analysis of catalogs of extragalactic objects. V.
  Three-point correlation function for the galaxy distribution in the Zwicky
  catalog.}}
\bjtitle{\apj}
\bvolume{196},
\bfpage{1}--\blpage{11}
(\byear{1975})
\doiurl{10.1086/153390}
\end{barticle}
\endbibitem

%%% 28
\bibitem[\protect\citeauthoryear{{Lanzetta} et~al.}{1995}]{lan1995}
\begin{barticle}
\bauthor{\bsnm{{Lanzetta}}, \binits{K.M.}},
\bauthor{\bsnm{{Bowen}}, \binits{D.V.}},
\bauthor{\bsnm{{Tytler}}, \binits{D.}},
\bauthor{\bsnm{{Webb}}, \binits{J.K.}}:
\batitle{{The Gaseous Extent of Galaxies and the Origin of Lyman-Alpha
  Absorption Systems: A Survey of Galaxies in the Fields of Hubble Space
  Telescope Spectroscopic Target QSOs}}.
\bjtitle{\apj}
\bvolume{442},
\bfpage{538}
(\byear{1995})
\doiurl{10.1086/175459}
\end{barticle}
\endbibitem

%%% 29
\bibitem[\protect\citeauthoryear{{Chen} et~al.}{2001}]{che2001}
\begin{barticle}
\bauthor{\bsnm{{Chen}}, \binits{H.-W.}},
\bauthor{\bsnm{{Lanzetta}}, \binits{K.M.}},
\bauthor{\bsnm{{Webb}}, \binits{J.K.}}:
\batitle{{The Origin of C IV Absorption Systems at Redshifts $z < 1$: Discovery
  of Extended C IV Envelopes around Galaxies}}.
\bjtitle{\apj}
\bvolume{556}(\bissue{1}),
\bfpage{158}--\blpage{163}
(\byear{2001})
\doiurl{10.1086/321537}
{\href{https://arxiv.org/abs/astro-ph/0104403}{{arXiv:astro-ph/0104403}}}
{[astro-ph]}
\end{barticle}
\endbibitem

%%% 30
\bibitem[\protect\citeauthoryear{{Bergeron} and {Boiss{\'e}}}{1991}]{ber1991}
\begin{barticle}
\bauthor{\bsnm{{Bergeron}}, \binits{J.}},
\bauthor{\bsnm{{Boiss{\'e}}}, \binits{P.}}:
\batitle{{A sample of galaxies giving rise to Mg II quasar absorption
  systems.}}
\bjtitle{\aap}
\bvolume{243},
\bfpage{344}
(\byear{1991})
\end{barticle}
\endbibitem

%%% 31
\bibitem[\protect\citeauthoryear{{Lanzetta} et~al.}{2024}]{lan2024a}
\begin{botherref}
\oauthor{\bsnm{{Lanzetta}}, \binits{K.M.}},
\oauthor{\bsnm{{Gromoll}}, \binits{S.}},
\oauthor{\bsnm{{Shara}}, \binits{M.M.}},
\oauthor{\bsnm{{Garland}}, \binits{D.} \bsuffix{James~{Valls{-}Gabaud}}},
\oauthor{\bsnm{{Walter}}, \binits{F.M.}},
\oauthor{\bsnm{{Webb}}, \binits{J.K.}}:
{Introducing the Condor Array Telescope. V. Deep Broad- and Narrow-Band Imaging
  Observations of the M81 Group}.
ApJS
(2024)
\end{botherref}
\endbibitem

%%% 32
\bibitem[\protect\citeauthoryear{{Wenger} et~al.}{2000}]{wen2000}
\begin{barticle}
\bauthor{\bsnm{{Wenger}}, \binits{M.}},
\bauthor{\bsnm{{Ochsenbein}}, \binits{F.}},
\bauthor{\bsnm{{Egret}}, \binits{D.}},
\bauthor{\bsnm{{Dubois}}, \binits{P.}},
\bauthor{\bsnm{{Bonnarel}}, \binits{F.}},
\bauthor{\bsnm{{Borde}}, \binits{S.}},
\bauthor{\bsnm{{Genova}}, \binits{F.}},
\bauthor{\bsnm{{Jasniewicz}}, \binits{G.}},
\bauthor{\bsnm{{Lalo{\"e}}}, \binits{S.}},
\bauthor{\bsnm{{Lesteven}}, \binits{S.}},
\bauthor{\bsnm{{Monier}}, \binits{R.}}:
\batitle{{The SIMBAD astronomical database. The CDS reference database for
  astronomical objects}}.
\bjtitle{A\&AS}
\bvolume{143},
\bfpage{9}--\blpage{22}
(\byear{2000})
\doiurl{10.1051/aas:2000332}
{\href{https://arxiv.org/abs/astro-ph/0002110}{{arXiv:astro-ph/0002110}}}
{[astro-ph]}
\end{barticle}
\endbibitem

\end{thebibliography}

\newpage

\begin{appendices}

\section*{Additional information}

\renewcommand{\figurename}{Extended Data Fig.}

\begin{figure}[h]
\centering
\includegraphics[width=0.50\linewidth, angle=0]{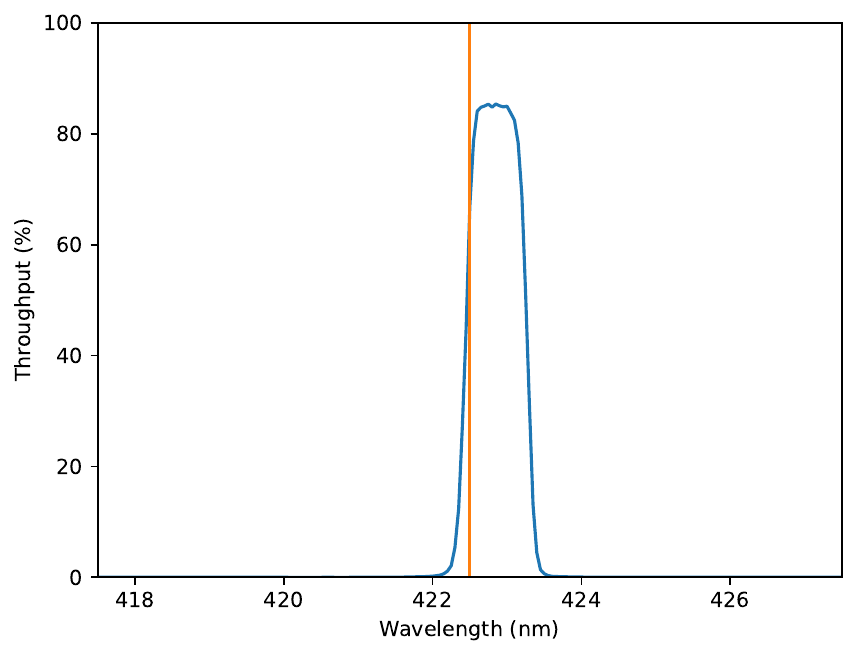}
\caption{Narrow-band filter throughput versus wavelength.  Blue curve shows
measured throughput (in a parallel beam), and orange line segment shows central
wavelength in Condor $f/5$ beam.  Central wavelength is $422.5$ nm, bandpass is
$1$ nm, and peak throughput is $\approx 85\%$.}
\end{figure}

\begin{landscape}
\begin{figure}
\centering
\vspace{-0.25in}
\centering
\includegraphics[width=0.90\linewidth, angle=0]{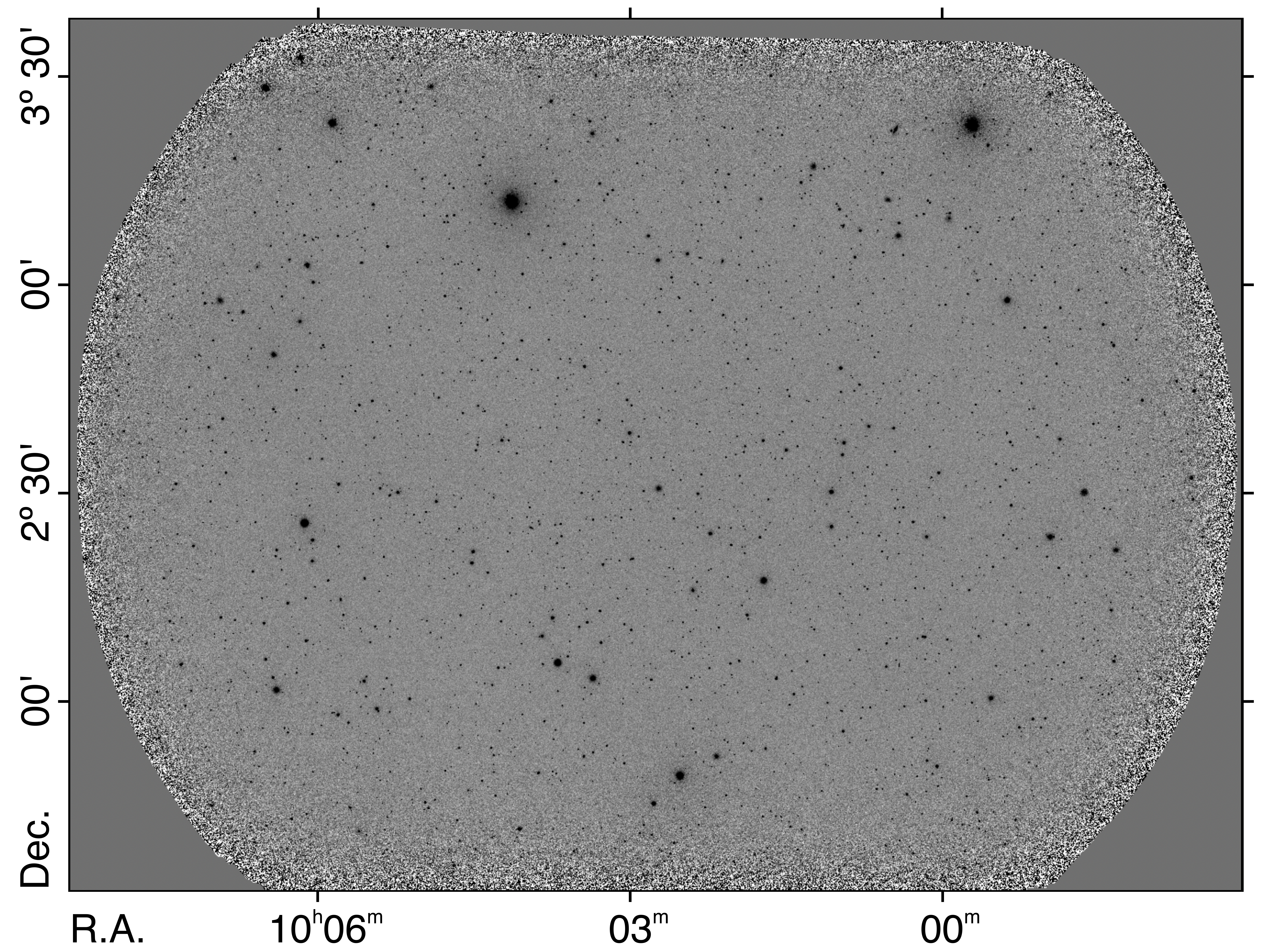}
\caption{Image of Condor field 416 through narrow-band filter of central
wavelength $422.5$ nm and bandpass $1$ nm.  Image spans $\approx 2.8 \times
1.3$ deg$^2$ at a pixel scale of $0.85$ arcsec pix$^{-1}$.  The COSMOS field
occupies the lower-right corner of the image.}
\end{figure}
\end{landscape}

\begin{landscape}
\begin{figure}
\centering
\vspace{-0.25in}
\centering
\includegraphics[width=0.90\linewidth, angle=0]{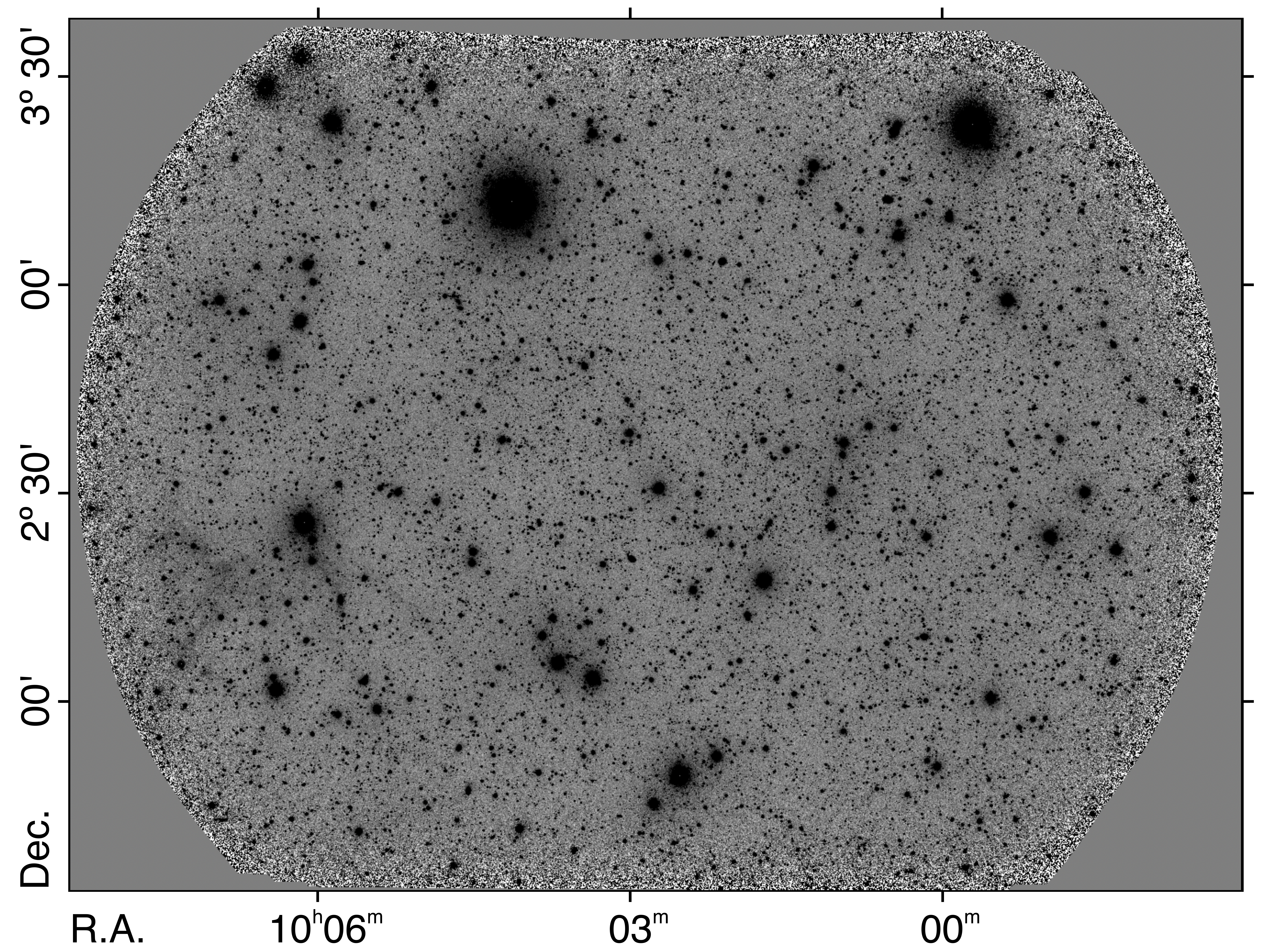}
\caption{Image of Condor field 416 through luminance filter.  Image spans
$\approx 2.8 \times 1.3$ deg$^2$ at a pixel scale of $0.85$ arcsec pix$^{-1}$.
The COSMOS field occupies the lower-right corner of the image.}
\end{figure}
\end{landscape}

\begin{landscape}
\begin{figure}
\centering
\vspace{-0.25in}
\centering
\includegraphics[width=0.90\linewidth, angle=0]{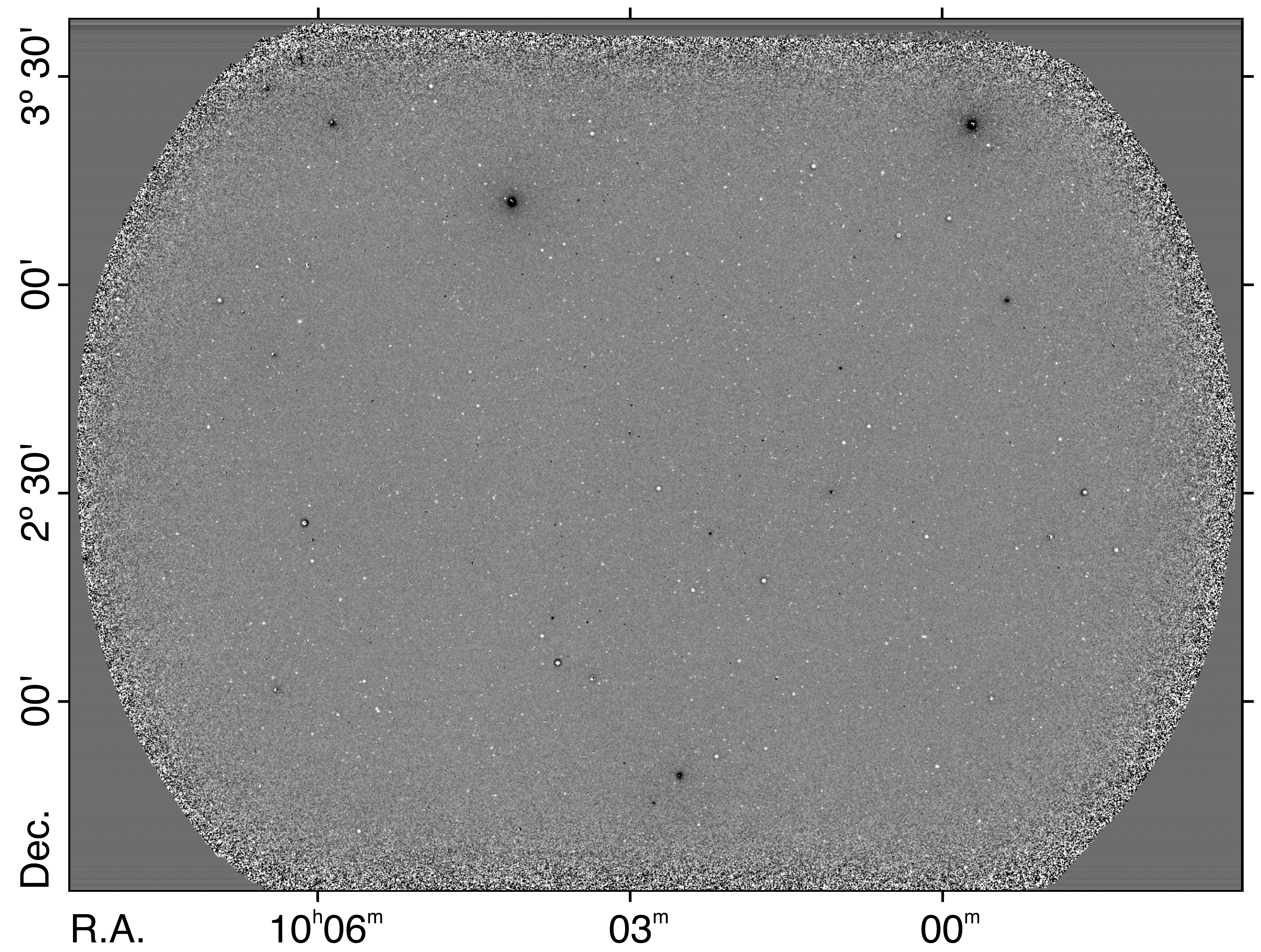}
\caption{Narrow-band minus luminance difference image of Condor field 416.
Image spans $\approx 2.8 \times 1.3$ deg$^2$ at a pixel scale of $0.85$ arcsec
pix$^{-1}$.  The COSMOS field occupies the lower-right corner of the image.}
\end{figure}
\end{landscape}

\begin{table}[h]
\centering
\hspace{-0.80in}
\begin{tabular}{p{1.00in}ccccccc}
\multicolumn{8}{c}{{\bf Extended Data Table 1:}  Catalog of Features} \\
\hline
\hline
& \multicolumn{2}{c}{J2000} & & $f$ & $\sigma_f$ & $r$ & $\langle x \rangle$ \\
\cline{2-3}
\cline{5-6}
\multicolumn{1}{c}{ID} & R.A. & Dec. & & \multicolumn{2}{c}{($10^{-18}$ erg s$^{-1}$ cm$^{-2}$)} & (kpc) & ($10^{-18}$ erg s$^{-1}$ cm$^{-2}$ arcsec$^{-2}$) \\
\hline
\multicolumn{1}{l}{1} \dotfill & 10:04:20.9 & +01:56:55.0 & & 7.1 & 1.7 & 14.9 & 0.39 \\
\multicolumn{1}{l}{2} \dotfill & 10:00:21.7 & +01:56:54.3 & & 8.5 & 1.7 & 14.6 & 0.49 \\
\multicolumn{1}{l}{3} \dotfill & 10:05:19.5 & +01:57:00.2 & & 3.4 & 0.7 & 6.0 & 1.16 \\
\multicolumn{1}{l}{4} \dotfill & 10:06:12.7 & +01:57:04.6 & & 15.2 & 3.1 & 26.8 & 0.26 \\
\multicolumn{1}{l}{5} \dotfill & 09:59:37.8 & +01:57:02.1 & & 6.4 & 1.0 & 8.9 & 0.98 \\
\multicolumn{1}{l}{6} \dotfill & 10:06:06.6 & +01:57:11.4 & & 8.1 & 1.7 & 14.9 & 0.45 \\
\multicolumn{1}{l}{7} \dotfill & 10:00:56.0 & +01:57:10.6 & & 3.4 & 0.7 & 6.0 & 1.19 \\
\multicolumn{1}{l}{8} \dotfill & 10:03:35.4 & +01:57:14.7 & & 11.7 & 2.4 & 20.8 & 0.33 \\
\multicolumn{1}{l}{9} \dotfill & 10:05:26.6 & +01:57:15.9 & & 1.7 & 0.3 & 3.0 & 2.34 \\
\multicolumn{1}{l}{10} \dotfill & 10:06:09.9 & +01:57:17.3 & & 5.0 & 1.0 & 8.9 & 0.77 \\
\multicolumn{1}{l}{11} \dotfill & 10:00:25.2 & +01:57:35.6 & & 17.8 & 3.1 & 26.8 & 0.30 \\
\multicolumn{1}{l}{12} \dotfill & 10:04:35.9 & +01:57:36.2 & & 3.6 & 0.7 & 6.0 & 1.26 \\
\multicolumn{1}{l}{13} \dotfill & 10:03:38.9 & +01:57:37.6 & & 8.0 & 1.7 & 14.9 & 0.44 \\
\multicolumn{1}{l}{14} \dotfill & 10:00:42.6 & +01:57:37.4 & & 8.1 & 1.7 & 14.9 & 0.45 \\
\multicolumn{1}{l}{15} \dotfill & 09:59:34.9 & +01:57:37.0 & & 9.4 & 1.7 & 14.9 & 0.52 \\
\multicolumn{1}{l}{16} \dotfill & 09:59:46.4 & +01:57:38.3 & & 7.2 & 1.4 & 11.9 & 0.62 \\
\multicolumn{1}{l}{17} \dotfill & 10:06:08.4 & +01:57:50.1 & & 22.8 & 5.4 & 47.6 & 0.12 \\
\multicolumn{1}{l}{18} \dotfill & 10:06:12.7 & +01:57:47.9 & & 5.7 & 1.0 & 8.9 & 0.87 \\
\multicolumn{1}{l}{19} \dotfill & 10:02:29.6 & +01:58:05.3 & & 12.0 & 2.0 & 17.8 & 0.46 \\
\multicolumn{1}{l}{20} \dotfill & 10:03:32.3 & +01:58:09.5 & & 6.4 & 1.4 & 11.9 & 0.56 \\
\multicolumn{1}{l}{21} \dotfill & 10:02:17.0 & +01:58:11.6 & & 9.2 & 1.7 & 14.9 & 0.51 \\
\multicolumn{1}{l}{22} \dotfill & 10:02:13.3 & +01:58:30.7 & & 12.2 & 2.0 & 17.8 & 0.47 \\
\multicolumn{1}{l}{23} \dotfill & 10:00:31.2 & +01:58:38.8 & & 15.6 & 2.8 & 24.4 & 0.32 \\
\multicolumn{1}{l}{24} \dotfill & 10:02:58.8 & +01:58:41.1 & & 14.9 & 2.4 & 21.2 & 0.40 \\
\multicolumn{1}{l}{25} \dotfill & 10:01:13.6 & +01:58:45.0 & & 23.7 & 5.4 & 47.6 & 0.13 \\
\multicolumn{1}{l}{26} \dotfill & 10:03:02.4 & +01:58:41.4 & & 10.3 & 1.7 & 14.9 & 0.57 \\
\multicolumn{1}{l}{27} \dotfill & 10:01:28.7 & +01:58:45.9 & & 10.0 & 2.0 & 17.8 & 0.38 \\
\multicolumn{1}{l}{28} \dotfill & 10:03:19.4 & +01:59:12.8 & & 5.3 & 1.0 & 8.9 & 0.82 \\
\multicolumn{1}{l}{29} \dotfill & 10:05:13.0 & +01:59:15.0 & & 8.2 & 1.7 & 14.9 & 0.45 \\
\multicolumn{1}{l}{30} \dotfill & 10:04:25.0 & +01:59:43.2 & & 10.1 & 2.4 & 20.8 & 0.29 \\
\multicolumn{1}{l}{31} \dotfill & 10:02:57.5 & +01:59:57.4 & & 10.6 & 2.0 & 17.8 & 0.41 \\
\multicolumn{1}{l}{32} \dotfill & 10:00:16.6 & +01:59:57.9 & & 6.9 & 1.4 & 11.9 & 0.60 \\
\multicolumn{1}{l}{33} \dotfill & 10:05:54.3 & +01:59:58.1 & & 5.1 & 1.0 & 8.9 & 0.79 \\
\multicolumn{1}{l}{34} \dotfill & 10:00:53.8 & +02:00:09.1 & & 16.3 & 3.3 & 29.1 & 0.24 \\
\multicolumn{1}{l}{35} \dotfill & 10:00:29.8 & +02:00:08.9 & & 21.9 & 3.2 & 28.1 & 0.34 \\
\multicolumn{1}{l}{36} \dotfill & 10:06:12.4 & +02:00:17.0 & & 10.9 & 2.0 & 17.8 & 0.42 \\
\multicolumn{1}{l}{37} \dotfill & 10:03:50.8 & +02:00:24.6 & & 3.2 & 0.7 & 6.0 & 1.10 \\
\multicolumn{1}{l}{38} \dotfill & 10:02:45.5 & +02:00:31.9 & & 16.6 & 3.1 & 26.8 & 0.28 \\
\multicolumn{1}{l}{39} \dotfill & 10:02:47.8 & +02:01:19.9 & & 7.3 & 1.4 & 11.9 & 0.63 \\
\multicolumn{1}{l}{40} \dotfill & 10:00:43.3 & +02:03:49.2 & & 19.6 & 4.1 & 35.7 & 0.19 \\
\multicolumn{1}{l}{41} \dotfill & 10:00:38.5 & +02:03:50.4 & & 16.7 & 3.1 & 26.8 & 0.29 \\
\multicolumn{1}{l}{42} \dotfill & 10:00:51.0 & +02:04:46.6 & & 37.8 & 7.1 & 62.5 & 0.12 \\
\multicolumn{1}{l}{43} \dotfill & 10:03:09.1 & +02:05:13.6 & & 10.2 & 2.0 & 17.8 & 0.39 \\
\multicolumn{1}{l}{44} \dotfill & 10:02:33.7 & +02:05:20.8 & & 8.7 & 1.7 & 14.9 & 0.48 \\
\multicolumn{1}{l}{45} \dotfill & 10:00:38.7 & +02:05:39.2 & & 34.5 & 6.2 & 54.2 & 0.14 \\
\multicolumn{1}{l}{46} \dotfill & 10:00:37.4 & +02:06:01.7 & & 38.8 & 6.7 & 58.9 & 0.14 \\
\multicolumn{1}{l}{47} \dotfill & 10:05:36.1 & +02:06:00.2 & & 5.1 & 1.0 & 8.9 & 0.78 \\
\multicolumn{1}{l}{48} \dotfill & 10:05:58.7 & +02:06:51.0 & & 5.1 & 1.0 & 8.9 & 0.78 \\
\multicolumn{1}{l}{49} \dotfill & 10:02:34.7 & +02:09:03.1 & & 12.0 & 2.0 & 17.8 & 0.46 \\
\multicolumn{1}{l}{50} \dotfill & 10:04:29.8 & +02:10:46.1 & & 5.3 & 1.0 & 8.9 & 0.82 \\
\multicolumn{1}{l}{51} \dotfill & 10:03:20.6 & +02:13:09.1 & & 5.4 & 1.0 & 8.9 & 0.84 \\
\multicolumn{1}{l}{52} \dotfill & 10:03:19.9 & +02:13:37.1 & & 19.4 & 2.7 & 24.0 & 0.41 \\
\multicolumn{1}{l}{53} \dotfill & 10:03:20.4 & +02:14:05.2 & & 4.9 & 1.0 & 8.9 & 0.76 \\
\multicolumn{1}{l}{54} \dotfill & 10:04:26.3 & +02:15:10.0 & & 3.4 & 0.7 & 6.0 & 1.17 \\
\multicolumn{1}{l}{55} \dotfill & 10:03:20.5 & +02:16:23.7 & & 8.4 & 1.7 & 14.9 & 0.47 \\
\multicolumn{1}{l}{56} \dotfill & 10:03:10.0 & +02:16:35.7 & & 15.8 & 3.1 & 26.8 & 0.27 \\
\multicolumn{1}{l}{57} \dotfill & 10:02:07.0 & +02:20:32.3 & & 7.3 & 1.4 & 11.9 & 0.63 \\
\multicolumn{1}{l}{58} \dotfill & 10:00:54.3 & +02:21:01.0 & & 14.6 & 2.7 & 23.8 & 0.31 \\
\multicolumn{1}{l}{59} \dotfill & 10:01:31.0 & +02:22:35.1 & & 5.4 & 1.0 & 8.9 & 0.84 \\
\multicolumn{1}{l}{60} \dotfill & 10:03:16.6 & +02:23:18.9 & & 7.5 & 1.4 & 11.9 & 0.64 \\
\multicolumn{1}{l}{61} \dotfill & 10:00:21.7 & +02:23:55.0 & & 3.7 & 0.7 & 6.0 & 1.29 \\
\multicolumn{1}{l}{62} \dotfill & 10:00:21.9 & +02:23:56.3 & & 2.7 & 0.3 & 3.0 & 3.74 \\
\multicolumn{1}{l}{63} \dotfill & 10:00:21.8 & +02:23:56.7 & & 5.3 & 0.7 & 6.0 & 1.84 \\
\multicolumn{1}{l}{64} \dotfill & 10:00:21.9 & +02:23:58.7 & & 7.9 & 1.1 & 9.4 & 1.09 \\
\hline
\end{tabular}
\end{table}

\begin{table}[h]
\centering
\hspace{-0.80in}
\begin{tabular}{p{1.00in}ccccccc}
\multicolumn{8}{c}{{\bf Extended Data Table 1:}  Catalog of Features} \\
\hline
\hline
& \multicolumn{2}{c}{J2000} & & $f$ & $\sigma_f$ & $r$ & $\langle x \rangle$ \\
\cline{2-3}
\cline{5-6}
\multicolumn{1}{c}{ID} & R.A. & Dec. & & \multicolumn{2}{c}{($10^{-18}$ erg s$^{-1}$ cm$^{-2}$)} & (kpc) & ($10^{-18}$ erg s$^{-1}$ cm$^{-2}$ arcsec$^{-2}$) \\
\hline
\multicolumn{1}{l}{65} \dotfill & 10:00:48.9 & +02:26:46.0 & & 7.8 & 1.4 & 11.9 & 0.67 \\
\multicolumn{1}{l}{66} \dotfill & 10:02:09.1 & +02:27:01.6 & & 17.6 & 2.7 & 23.8 & 0.38 \\
\multicolumn{1}{l}{67} \dotfill & 10:05:13.7 & +02:28:44.9 & & 8.0 & 1.4 & 11.9 & 0.69 \\
\multicolumn{1}{l}{68} \dotfill & 10:04:05.5 & +02:32:22.0 & & 15.8 & 3.4 & 29.8 & 0.22 \\
\multicolumn{1}{l}{69} \dotfill & 10:01:16.5 & +02:32:29.1 & & 5.8 & 1.0 & 8.9 & 0.89 \\
\multicolumn{1}{l}{70} \dotfill & 10:02:01.1 & +02:36:17.5 & & 7.8 & 1.4 & 11.9 & 0.67 \\
\multicolumn{1}{l}{71} \dotfill & 10:01:51.6 & +02:36:17.1 & & 1.8 & 0.3 & 3.0 & 2.47 \\
\multicolumn{1}{l}{72} \dotfill & 10:01:20.9 & +02:36:18.2 & & 4.7 & 0.7 & 6.0 & 1.62 \\
\multicolumn{1}{l}{73} \dotfill & 10:04:50.2 & +02:39:15.7 & & 15.7 & 2.7 & 23.8 & 0.34 \\
\multicolumn{1}{l}{74} \dotfill & 10:03:45.2 & +02:40:24.4 & & 8.1 & 1.7 & 14.9 & 0.45 \\
\multicolumn{1}{l}{75} \dotfill & 10:01:01.0 & +02:40:59.0 & & 5.6 & 1.0 & 8.9 & 0.86 \\
\multicolumn{1}{l}{76} \dotfill & 10:03:04.8 & +02:41:09.9 & & 7.0 & 1.4 & 11.9 & 0.61 \\
\multicolumn{1}{l}{77} \dotfill & 10:00:37.6 & +02:41:42.7 & & 3.6 & 0.7 & 6.0 & 1.26 \\
\multicolumn{1}{l}{78} \dotfill & 10:05:01.9 & +02:43:25.5 & & 22.1 & 4.1 & 35.7 & 0.21 \\
\multicolumn{1}{l}{79} \dotfill & 09:59:54.0 & +02:44:48.0 & & 5.7 & 1.0 & 8.9 & 0.87 \\
\multicolumn{1}{l}{80} \dotfill & 10:01:35.1 & +02:45:26.9 & & 5.3 & 1.0 & 8.9 & 0.81 \\
\multicolumn{1}{l}{81} \dotfill & 10:05:56.8 & +02:46:36.7 & & 15.8 & 2.4 & 20.8 & 0.45 \\
\multicolumn{1}{l}{82} \dotfill & 10:03:16.6 & +02:48:54.5 & & 8.5 & 1.7 & 14.9 & 0.47 \\
\multicolumn{1}{l}{83} \dotfill & 10:05:57.9 & +02:49:24.5 & & 12.6 & 2.0 & 17.8 & 0.49 \\
\multicolumn{1}{l}{84} \dotfill & 10:00:06.1 & +02:49:41.0 & & 15.2 & 2.7 & 23.8 & 0.33 \\
\multicolumn{1}{l}{85} \dotfill & 10:01:41.5 & +02:51:57.1 & & 49.5 & 8.5 & 74.4 & 0.11 \\
\multicolumn{1}{l}{86} \dotfill & 10:01:50.5 & +02:53:57.8 & & 13.4 & 2.4 & 20.8 & 0.38 \\
\multicolumn{1}{l}{87} \dotfill & 10:01:01.3 & +02:56:50.2 & & 12.4 & 2.4 & 20.8 & 0.35 \\
\multicolumn{1}{l}{88} \dotfill & 10:01:25.7 & +02:58:09.8 & & 7.1 & 1.4 & 11.9 & 0.61 \\
\multicolumn{1}{l}{89} \dotfill & 10:01:34.4 & +02:59:11.8 & & 11.6 & 2.0 & 17.8 & 0.45 \\
\multicolumn{1}{l}{90} \dotfill & 10:02:04.1 & +03:00:59.0 & & 17.8 & 3.4 & 29.8 & 0.25 \\
\multicolumn{1}{l}{91} \dotfill & 10:00:45.5 & +03:02:13.9 & & 5.1 & 1.0 & 8.9 & 0.79 \\
\multicolumn{1}{l}{92} \dotfill & 10:03:37.4 & +03:02:57.6 & & 5.8 & 1.0 & 8.9 & 0.89 \\
\multicolumn{1}{l}{93} \dotfill & 10:00:41.3 & +03:03:19.3 & & 34.5 & 5.2 & 45.4 & 0.20 \\
\multicolumn{1}{l}{94} \dotfill & 10:01:45.3 & +03:05:07.6 & & 5.2 & 1.0 & 8.9 & 0.79 \\
\multicolumn{1}{l}{95} \dotfill & 09:59:49.2 & +03:05:32.3 & & 5.7 & 1.0 & 8.9 & 0.88 \\
\multicolumn{1}{l}{96} \dotfill & 10:02:09.8 & +03:06:11.4 & & 17.3 & 3.1 & 26.8 & 0.30 \\
\multicolumn{1}{l}{97} \dotfill & 10:06:11.7 & +03:06:25.5 & & 5.4 & 1.0 & 8.9 & 0.83 \\
\multicolumn{1}{l}{98} \dotfill & 10:06:01.4 & +03:07:12.8 & & 7.5 & 1.4 & 11.9 & 0.65 \\
\multicolumn{1}{l}{99} \dotfill & 09:59:39.7 & +03:08:12.8 & & 1.8 & 0.3 & 3.0 & 2.47 \\
\multicolumn{1}{l}{100} \dotfill & 10:02:44.2 & +03:08:17.7 & & 3.5 & 0.7 & 6.0 & 1.23 \\
\multicolumn{1}{l}{101} \dotfill & 10:05:59.7 & +03:09:17.8 & & 12.2 & 2.0 & 17.8 & 0.47 \\
\multicolumn{1}{l}{102} \dotfill & 10:02:11.8 & +03:09:39.2 & & 3.4 & 0.7 & 6.0 & 1.18 \\
\multicolumn{1}{l}{103} \dotfill & 10:06:00.2 & +03:09:53.9 & & 1.7 & 0.3 & 3.0 & 2.38 \\
\multicolumn{1}{l}{104} \dotfill & 09:59:42.8 & +03:09:56.1 & & 7.9 & 1.4 & 11.9 & 0.69 \\
\multicolumn{1}{l}{105} \dotfill & 10:06:15.9 & +03:10:35.3 & & 5.8 & 1.0 & 8.9 & 0.89 \\
\multicolumn{1}{l}{106} \dotfill & 10:06:00.6 & +03:10:56.8 & & 1.7 & 0.3 & 3.0 & 2.38 \\
\multicolumn{1}{l}{107} \dotfill & 10:01:46.6 & +03:11:01.2 & & 11.7 & 2.4 & 20.8 & 0.33 \\
\multicolumn{1}{l}{108} \dotfill & 10:06:15.3 & +03:11:11.0 & & 5.8 & 1.0 & 8.9 & 0.89 \\
\multicolumn{1}{l}{109} \dotfill & 10:02:36.2 & +03:11:25.1 & & 5.7 & 1.0 & 8.9 & 0.88 \\
\multicolumn{1}{l}{110} \dotfill & 10:04:47.3 & +03:11:32.3 & & 5.5 & 1.0 & 8.9 & 0.84 \\
\multicolumn{1}{l}{111} \dotfill & 10:04:47.0 & +03:11:42.2 & & 9.8 & 1.5 & 13.0 & 0.72 \\
\multicolumn{1}{l}{112} \dotfill & 10:01:06.1 & +03:11:41.4 & & 3.6 & 0.7 & 6.0 & 1.24 \\
\hline
\end{tabular}
\end{table}

\renewcommand{\thefigure}{5}
\begin{landscape}
\begin{figure}
\centering
\vspace{-0.25in}
\includegraphics[width=1.00\linewidth, angle=0, viewport=97 62 745 530]{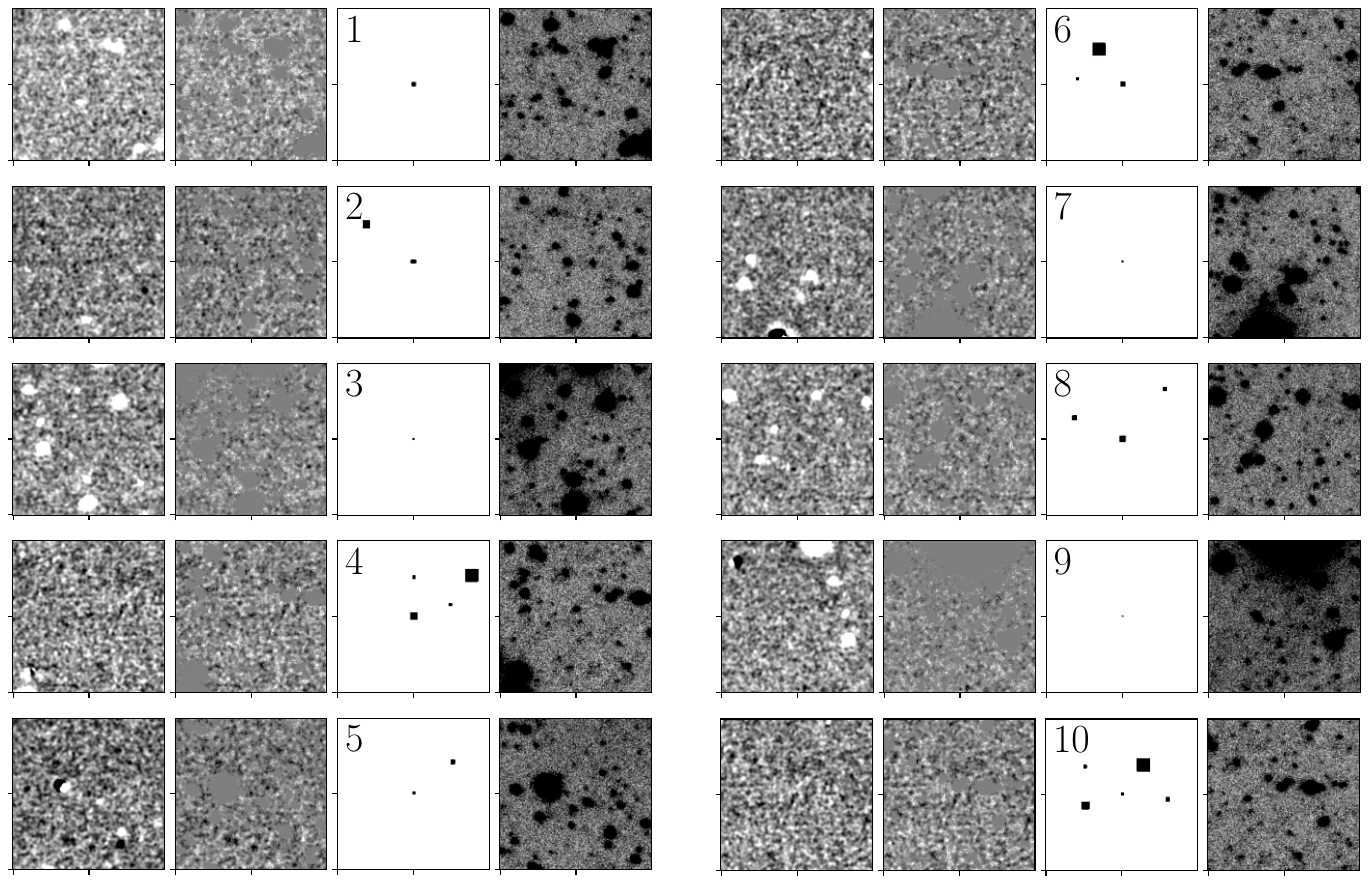}
\caption{Detected features of (1) high statistical significance and (2) large
rest-frame equivalent width (for \lya).  Each panel shows small portions of
(left to right) difference image, masked difference image, detection map, and
luminance image.  Images extend $2.8 \times 2.8$ armin$^2$, which at redshift
$z = 2.4754$ corresponds to $1.4$ proper Mpc or $4.9$ Mpc comoving Mpc on a
side.  North is up and east is to the left.}
\end{figure}
\end{landscape}

\renewcommand{\thefigure}{5 (continued)}
\begin{landscape}
\begin{figure}
\centering
\vspace{-0.25in}
\includegraphics[width=1.00\linewidth, angle=0, viewport=97 62 745 530]{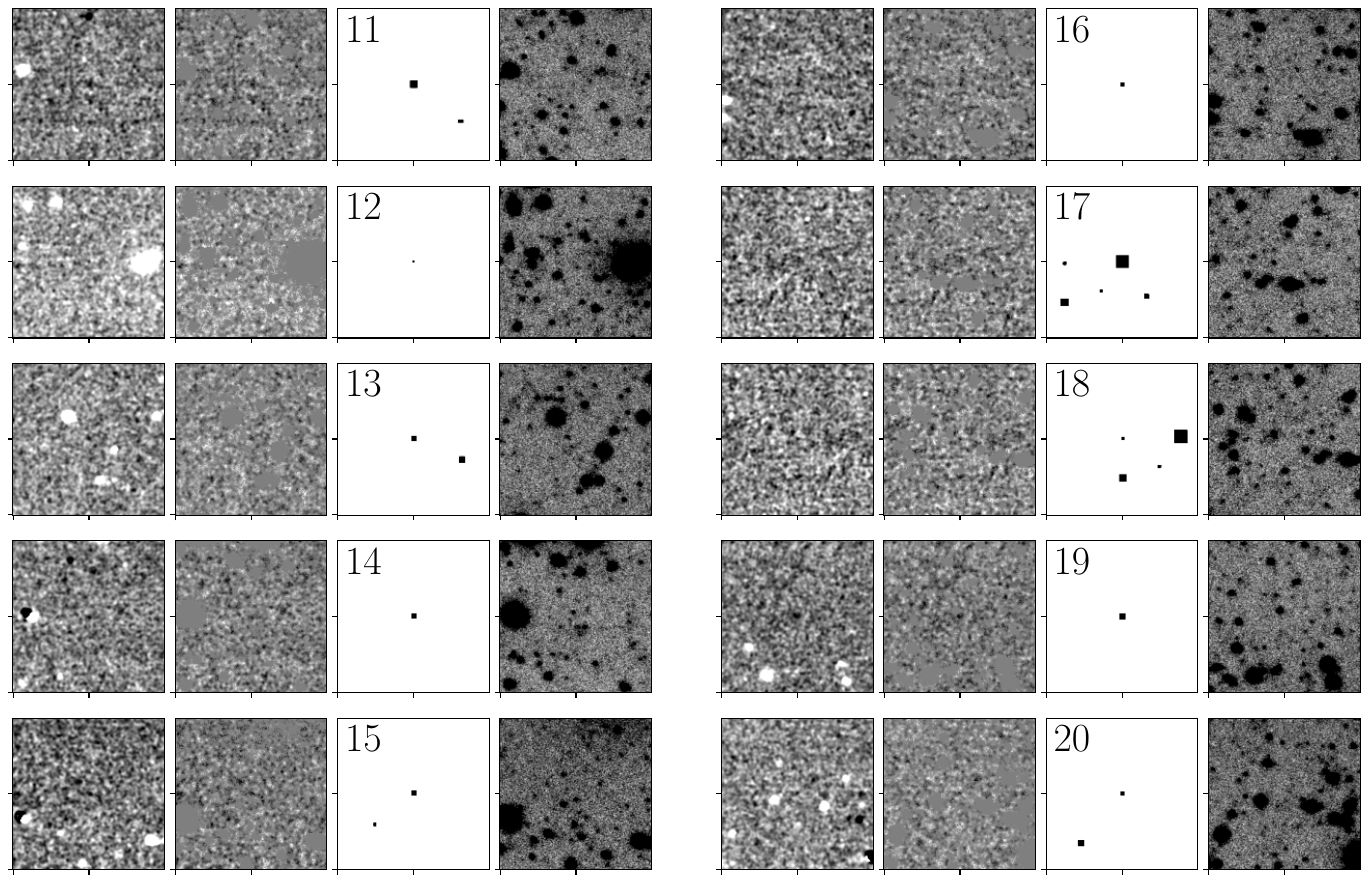}
\caption{}
\end{figure}
\end{landscape}

\renewcommand{\thefigure}{5 (continued)}
\begin{landscape}
\begin{figure}
\centering
\vspace{-0.25in}
\includegraphics[width=1.00\linewidth, angle=0, viewport=97 62 745 530]{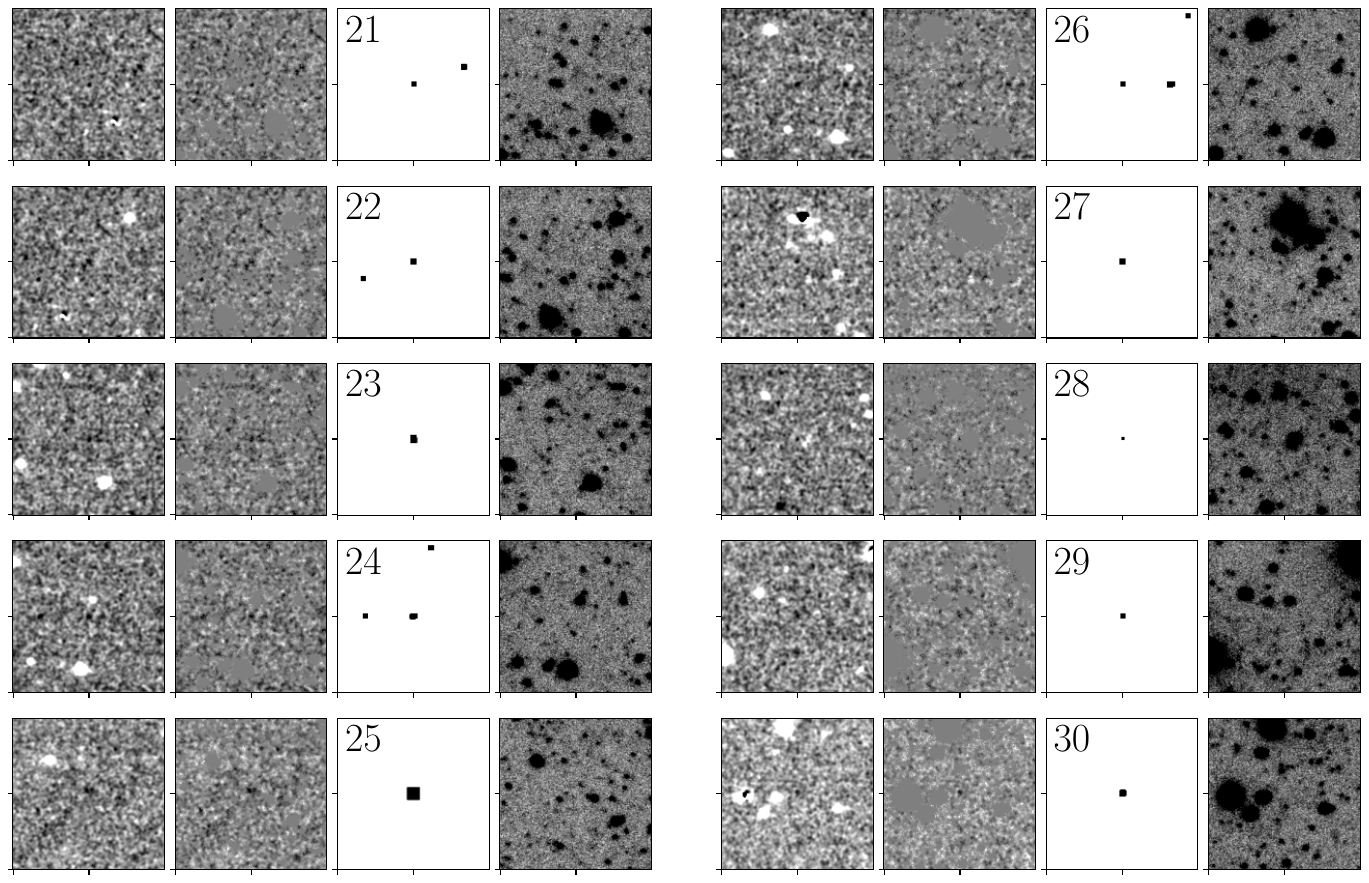}
\caption{}
\end{figure}
\end{landscape}

\renewcommand{\thefigure}{5 (continued)}
\begin{landscape}
\begin{figure}
\centering
\vspace{-0.25in}
\includegraphics[width=1.00\linewidth, angle=0, viewport=97 62 745 530]{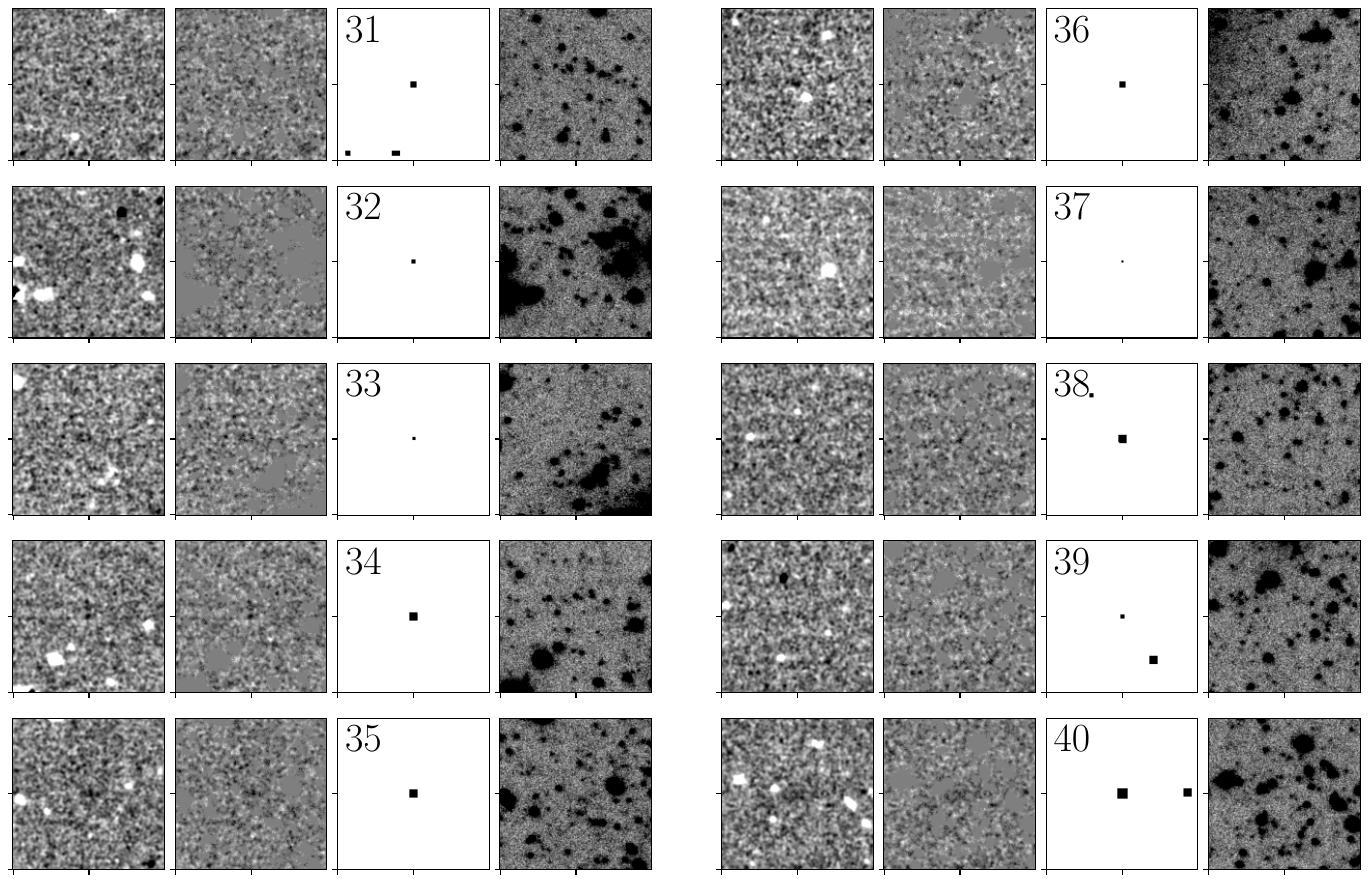}
\caption{}
\end{figure}
\end{landscape}

\renewcommand{\thefigure}{5 (continued)}
\begin{landscape}
\begin{figure}
\centering
\vspace{-0.25in}
\includegraphics[width=1.00\linewidth, angle=0, viewport=97 62 745 530]{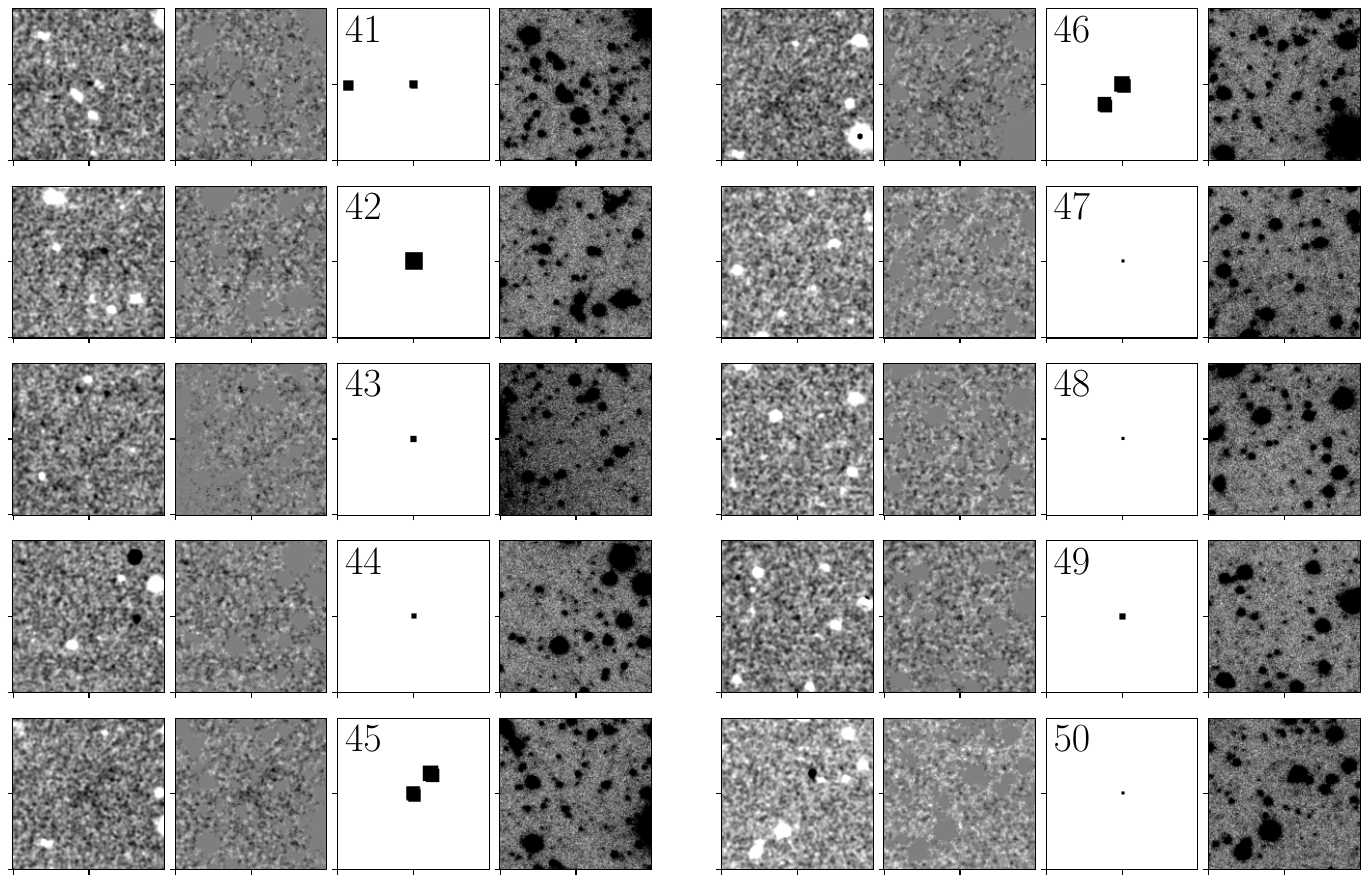}
\caption{}
\end{figure}
\end{landscape}

\renewcommand{\thefigure}{5 (continued)}
\begin{landscape}
\begin{figure}
\centering
\vspace{-0.25in}
\includegraphics[width=1.00\linewidth, angle=0, viewport=97 62 745 530]{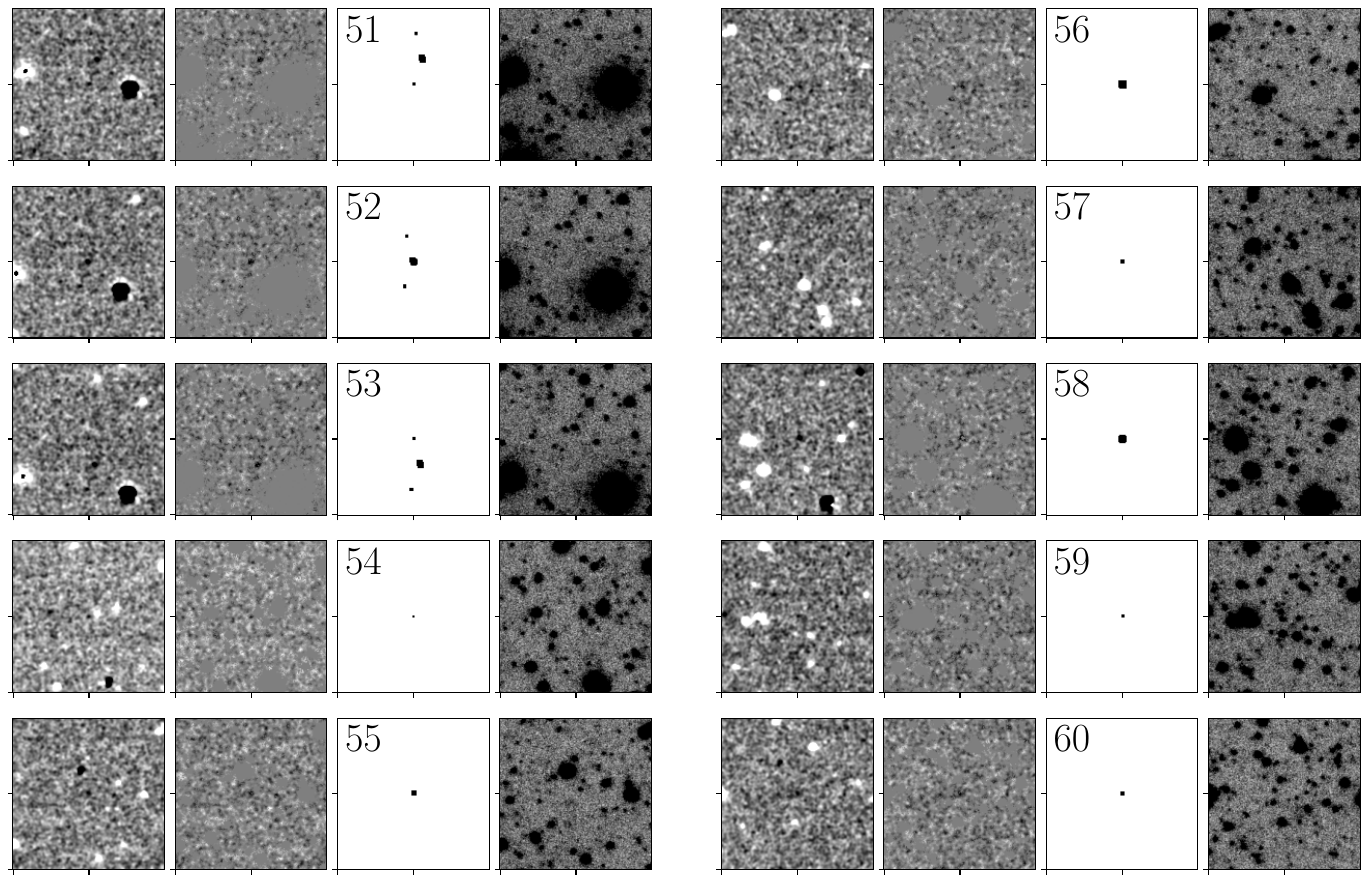}
\caption{}
\end{figure}
\end{landscape}

\renewcommand{\thefigure}{5 (continued)}
\begin{landscape}
\begin{figure}
\centering
\vspace{-0.25in}
\includegraphics[width=1.00\linewidth, angle=0, viewport=97 62 745 530]{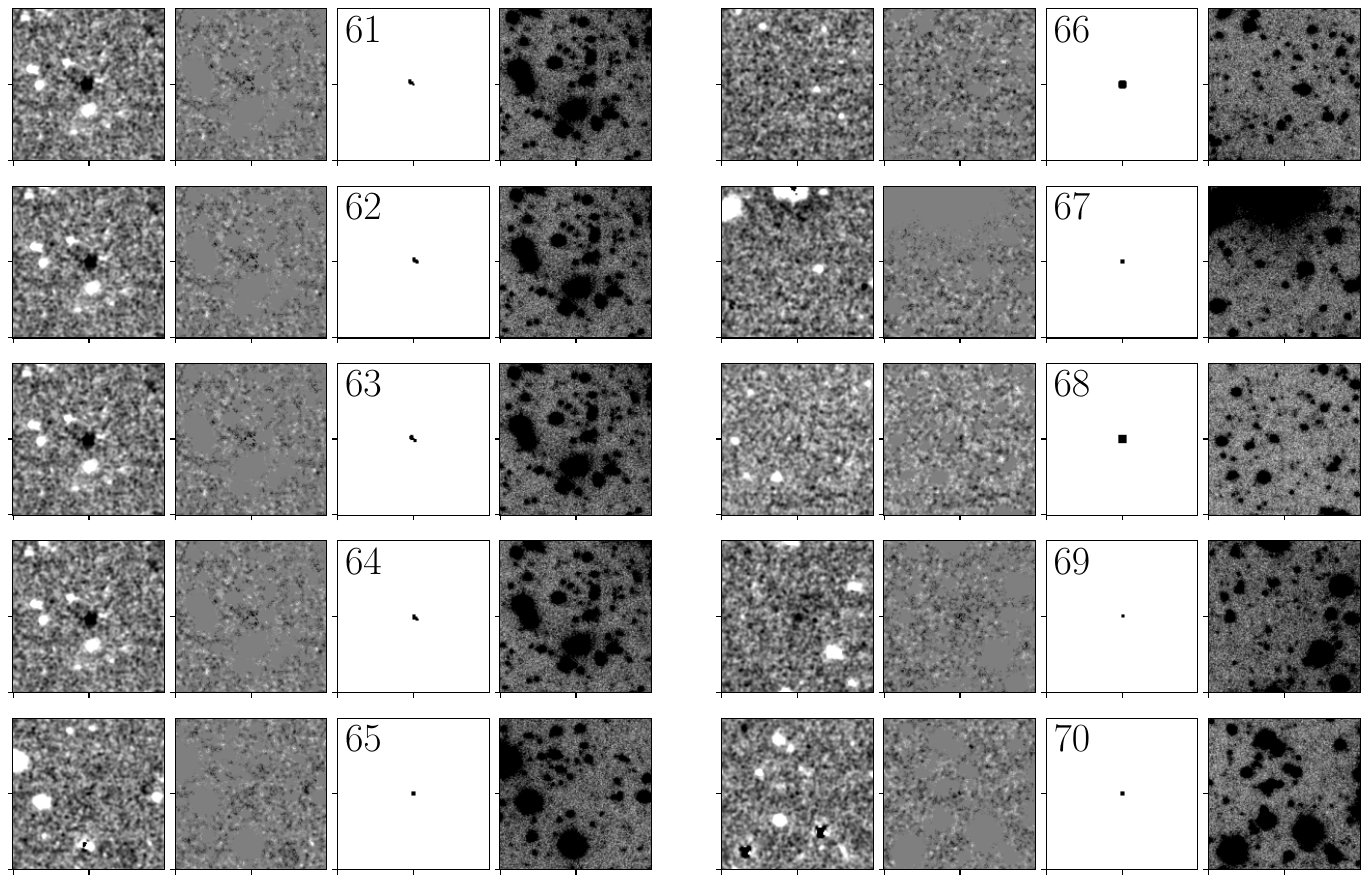}
\caption{}
\end{figure}
\end{landscape}

\renewcommand{\thefigure}{5 (continued)}
\begin{landscape}
\begin{figure}
\centering
\vspace{-0.25in}
\includegraphics[width=1.00\linewidth, angle=0, viewport=97 62 745 530]{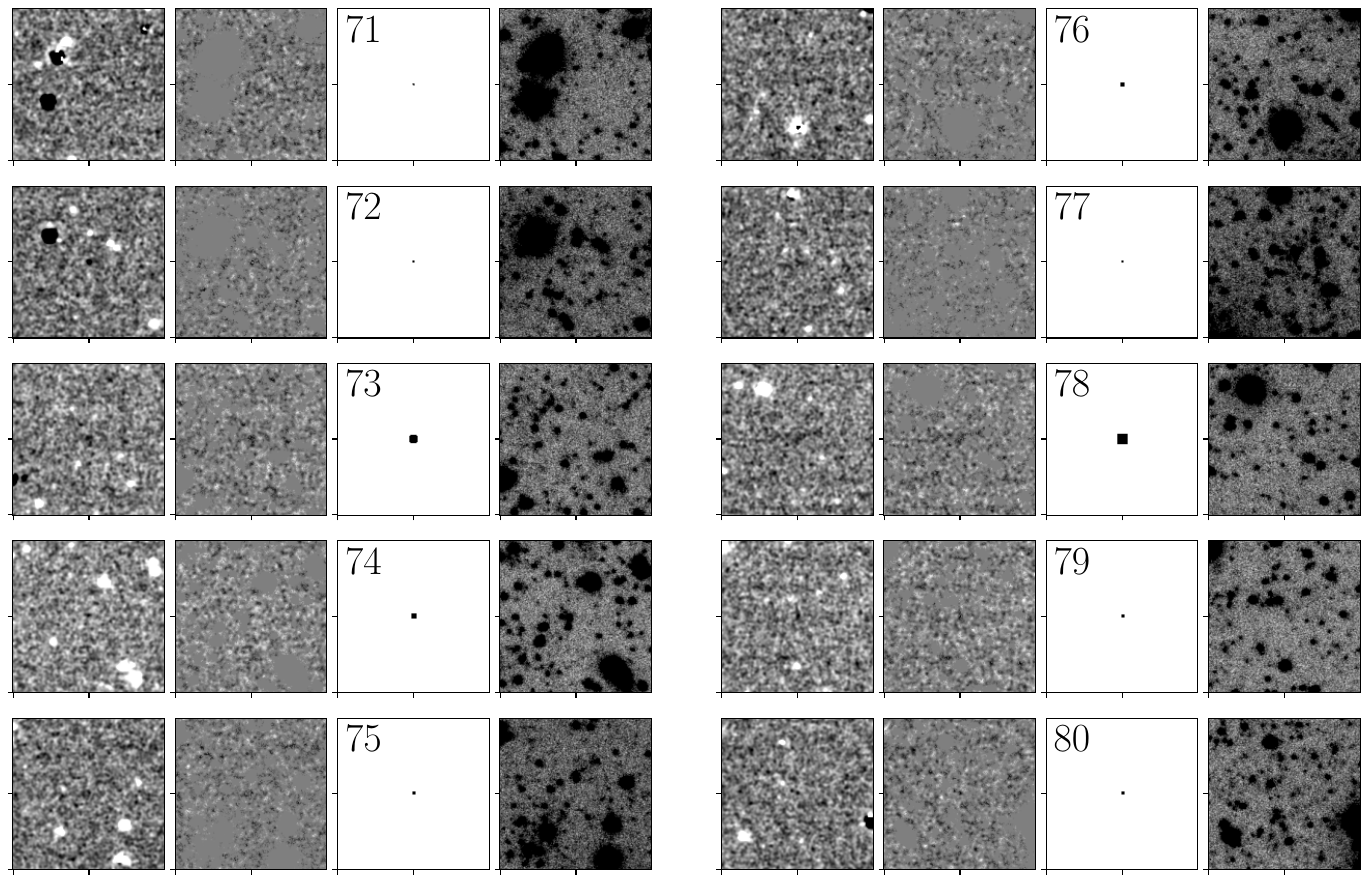}
\caption{}
\end{figure}
\end{landscape}

\renewcommand{\thefigure}{5 (continued)}
\begin{landscape}
\begin{figure}
\centering
\vspace{-0.25in}
\includegraphics[width=1.00\linewidth, angle=0, viewport=97 62 745 530]{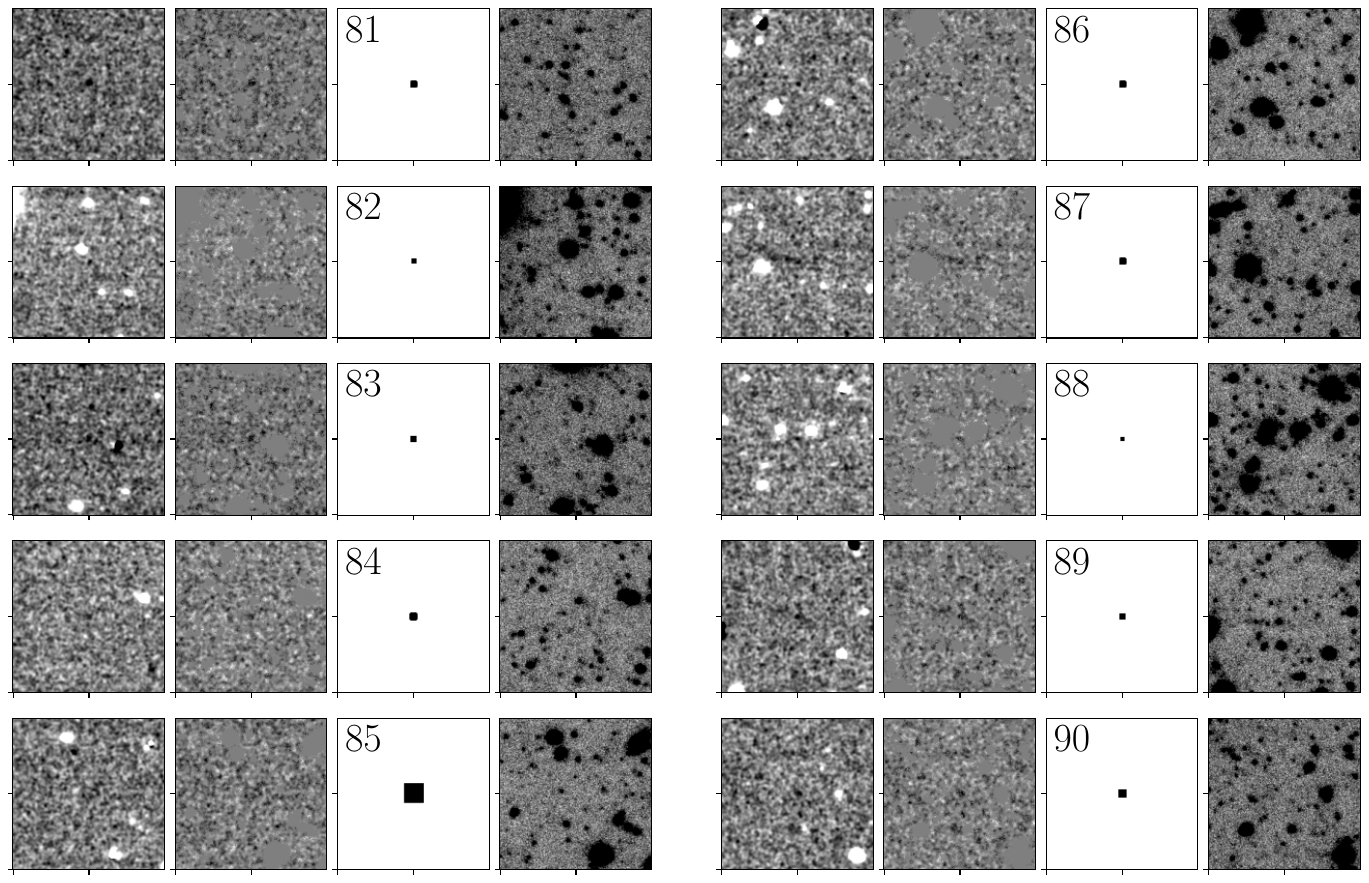}
\caption{}
\end{figure}
\end{landscape}

\renewcommand{\thefigure}{5 (continued)}
\begin{landscape}
\begin{figure}
\centering
\vspace{-0.25in}
\includegraphics[width=1.00\linewidth, angle=0, viewport=97 62 745 530]{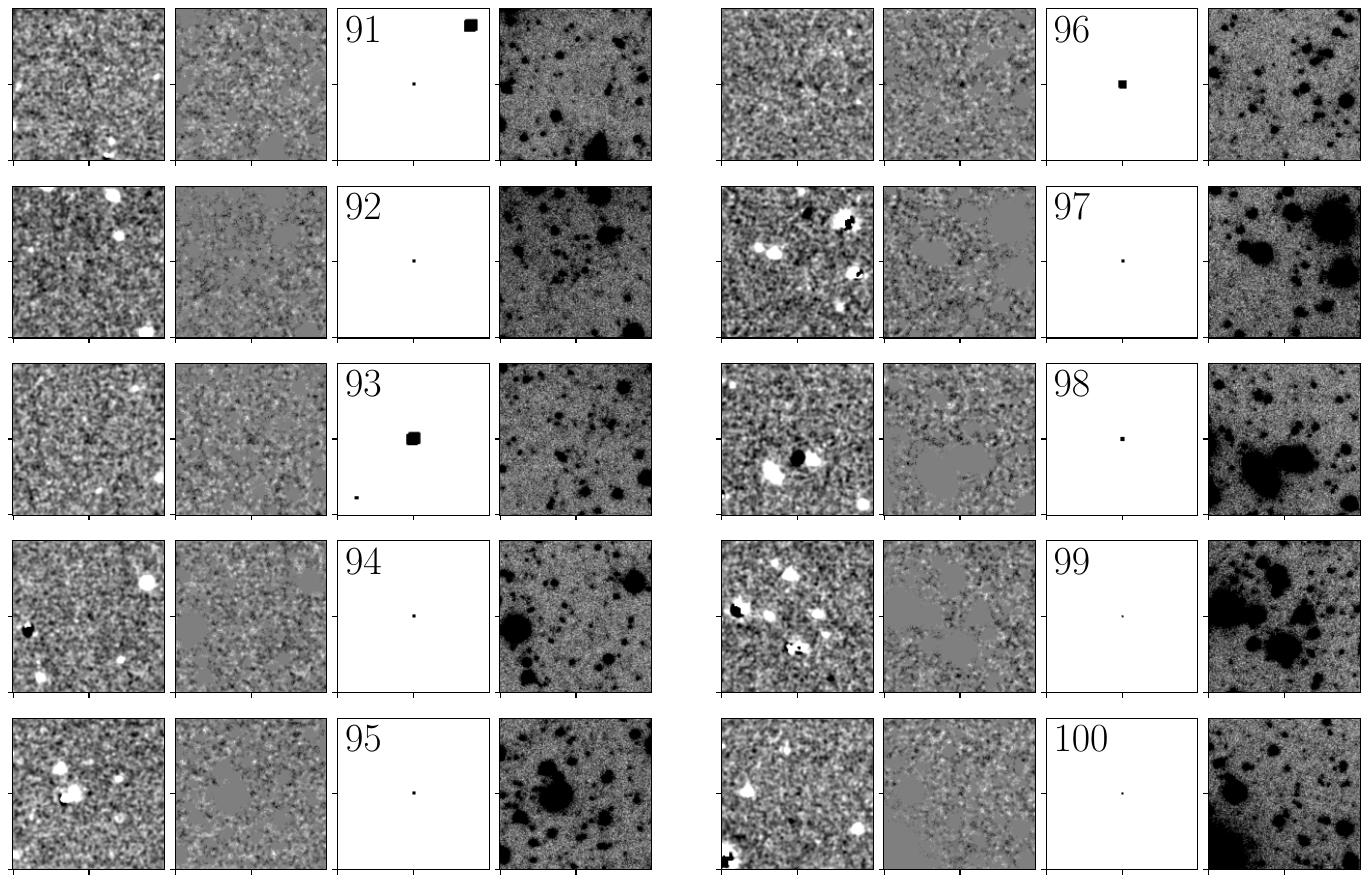}
\caption{}
\end{figure}
\end{landscape}

\renewcommand{\thefigure}{5 (continued)}
\begin{landscape}
\begin{figure}
\centering
\vspace{-0.25in}
\includegraphics[width=1.00\linewidth, angle=0, viewport=97 62 745 530]{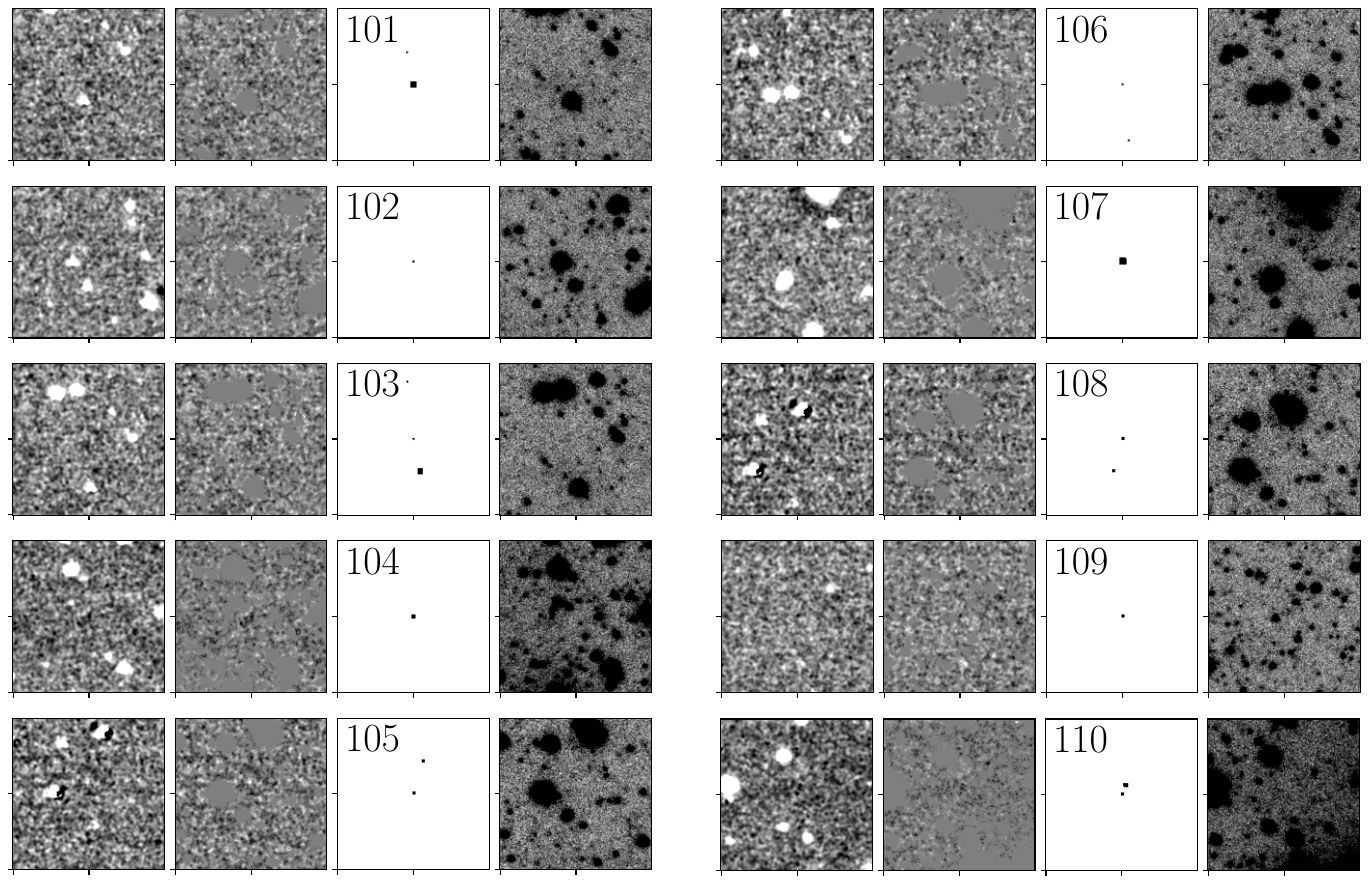}
\caption{}
\end{figure}
\end{landscape}

\renewcommand{\thefigure}{5 (continued)}
\begin{landscape}
\begin{figure}
\centering
\vspace{-0.25in}
\includegraphics[width=1.00\linewidth, angle=0, viewport=97 62 745 530]{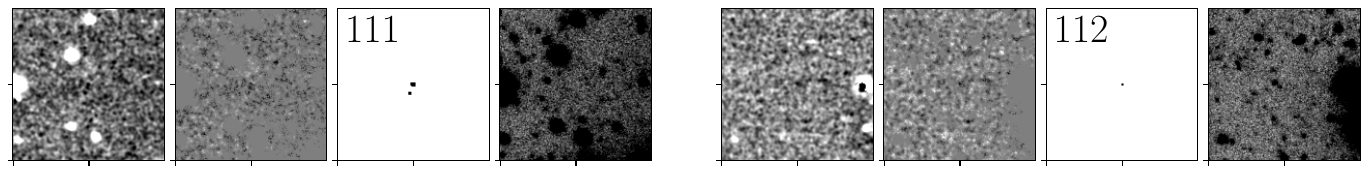}
\caption{}
\end{figure}
\end{landscape}

\renewcommand{\thefigure}{6}
\begin{figure}
\centering
\subfloat{
  \includegraphics[width=0.33\linewidth, angle=0]{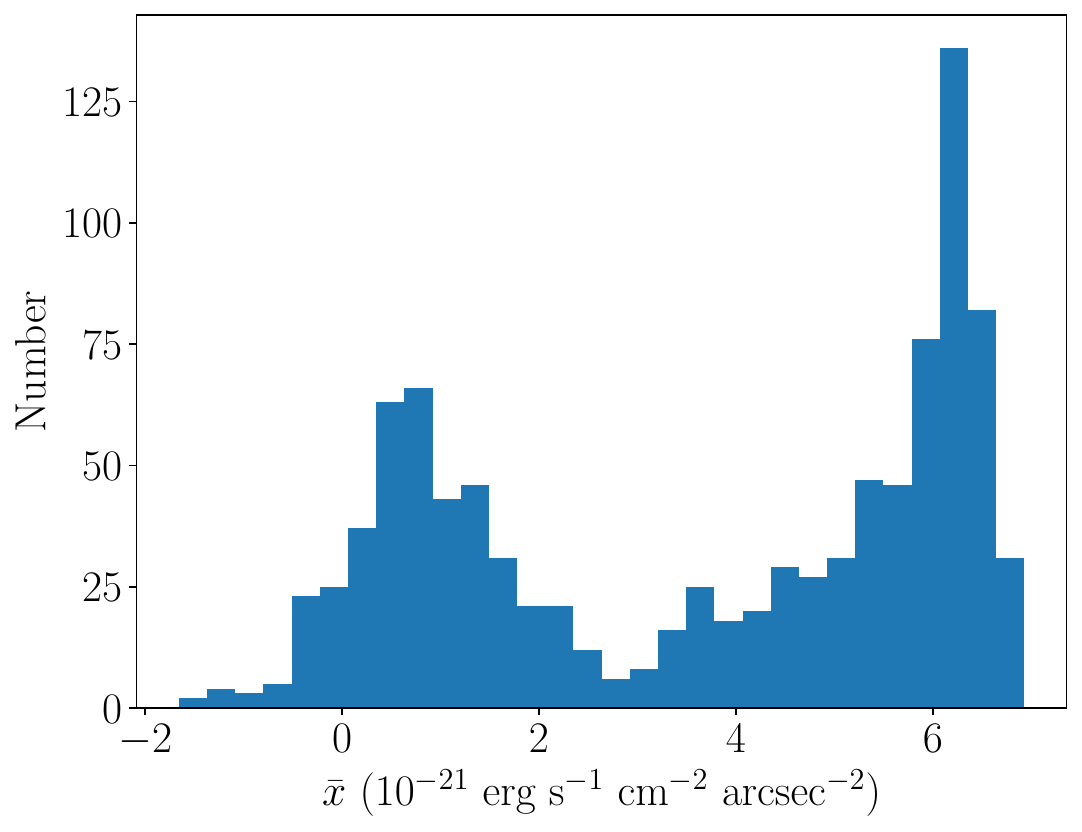}
}
\subfloat{
  \includegraphics[width=0.33\linewidth, angle=0]{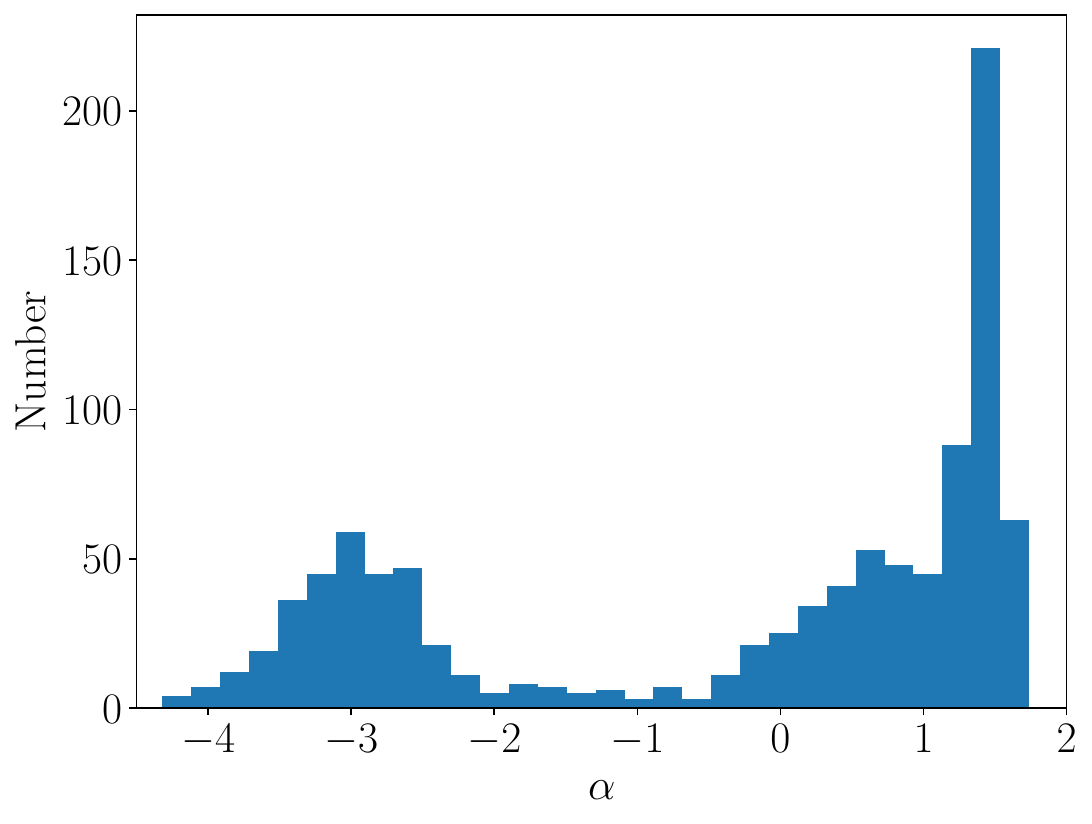}
}
\subfloat{
  \includegraphics[width=0.33\linewidth, angle=0]{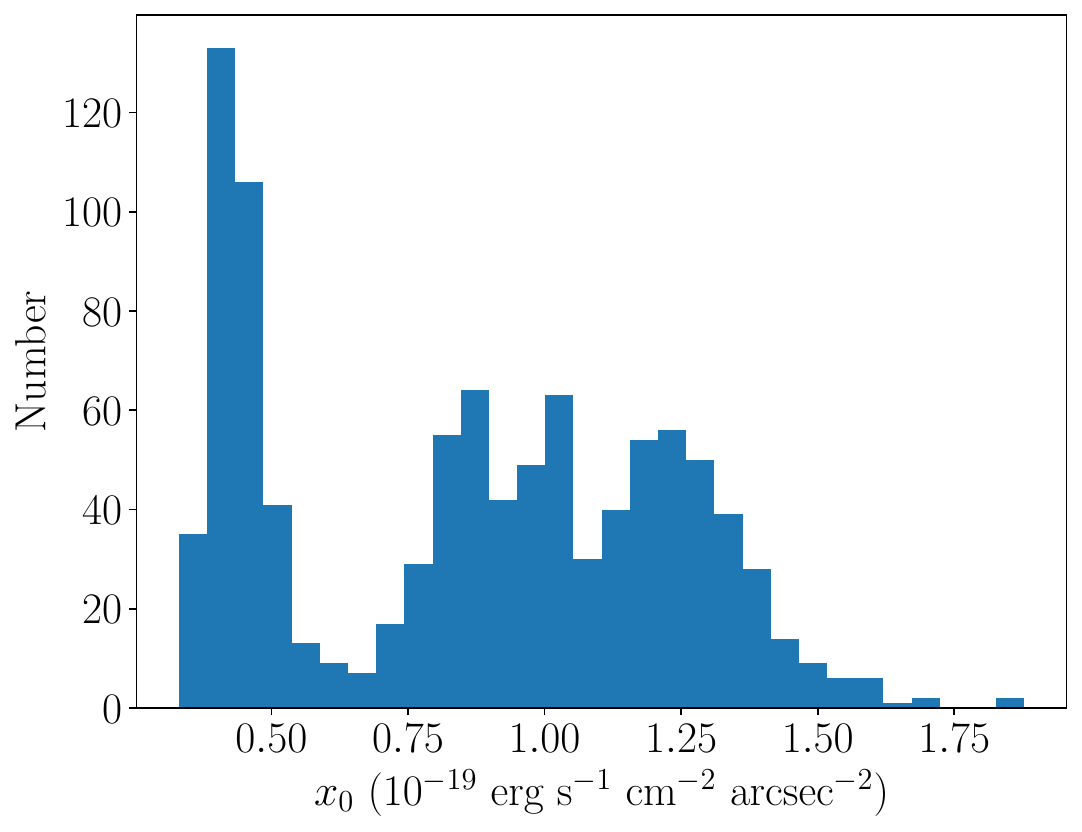}
}
\caption{Distributions of parameter values obtained in 1000 resamplings for 
parameters $\bar{x}$, $\alpha$, and $x_0$ of model of equation (2).
 Distributions of parameter values are bimodal, with values given in caption of
Fig.\ 2 occurring in right-most peak in all cases.}
\end{figure}

\end{appendices}

\end{document}